\documentclass[format=acmsmall,nonacm,screen=true,review=false]{acmart}

\usepackage[utf8]{inputenc}
\usepackage{include_packages}

\acmConference{}{}{} 
\startPage{1} 
\AtBeginDocument{
\fancypagestyle{mystyle}{
	\fancyhf{}%
	\fancyhead[RE]{\footnotesize\shortauthors}%
    \fancyhead[LO]{\footnotesize\shorttitle}%
	\fancyhead[RO,LE]{\footnotesize\thepage}%
}
\pagestyle{mystyle}
}

\setcitestyle{acmnumeric}

\title[When Smoothness is Not Enough]{When Smoothness is Not Enough: Toward Exact Quantification and Optimization of the Price of Anarchy}
\author{Rahul Chandan}
\affiliation{University of California, Santa Barbara}
\email{rchandan@ucsb.edu}
\author{Dario Paccagnan}
\affiliation{Imperial College London}
\email{d.paccagnan@imperial.ac.uk}
\author{Jason R. Marden}
\affiliation{University of California, Santa Barbara}
\email{jrmarden@ece.ucsb.edu}
\authorsaddresses{Authors' addresses: R.~Chandan, Department of Electrical and Computer Engineering, University of California, Santa Barbara (e-mail: \href{mailto:rchandan@ucsb.edu}{rchandan@ucsb.edu}); D.~Paccagnan, Department of Computing, Imperial College London (e-mail: \href{mailto:d.paccagnan@imperial.ac.uk}{d.paccagnan@imperial.ac.uk}); J.~R.~Marden, Department of Electrical and Computer Engineering, University of California, Santa Barbara (e-mail: \href{mailto:jrmarden@ece.ucsb.edu}{jrmarden@ece.ucsb.edu}).}

\begin{abstract}%
    The price of anarchy (PoA) is a popular metric for analyzing the inefficiency of self-interested decision making. Although its study is widespread, characterizing the PoA can be challenging. A commonly employed approach is based on the {\it smoothness framework}, which provides tight PoA values under the assumption that the system objective consists in the sum of the agents' individual welfares. Unfortunately, several important classes of problems do not satisfy this requirement (e.g., taxation in congestion games), and our first result demonstrates that the smoothness framework does \emph{not} tightly characterize the PoA for such settings. Motivated by this observation, this work develops a framework that achieves two chief objectives: i) to tightly characterize the PoA for such scenarios, and ii) to do so through a tractable approach. As a direct consequence, the proposed framework recovers and generalizes many existing PoA results, and enables efficient computation of incentives that optimize the PoA. We conclude by highlighting the applicability of our contributions to incentive design in congestion games and utility design in distributed welfare games.%
\end{abstract} 
\thanks{For the interested reader, the authors provide a software package, available in both MATLAB\textsuperscript{\textregistered} and Python, 
that implements the techniques described in this manuscript 
at~\href{https://github.com/rahul-chandan/resalloc-poa}{https://github.com/rahul-chandan/resalloc-poa}.}


\begin{document}

    \maketitle
    \renewcommand{\shortauthors}{R. Chandan, D. Paccagnan and J. R. Marden}


\section{Introduction}

The widespread proliferation of smartphones and other smart devices has led to a 
momentous shift in the operation of shared technological infrastructure like 
road-traffic networks, cloud computing and the power grid, where the local 
behaviours and interactions of individual decision makers are increasingly 
influencing the system wide performance.
Although such performance could be improved if a central coordinator 
was able to dictate the choices of individual decision makers, this approach is 
often infeasible owing to the distributed and self-interested nature of the very 
same decision making process.
Within these settings, wide ranging inefficiencies can severely degrade system 
performance, a phenomenon that is typically referred to as the \emph{tragedy 
of the commons}~\cite{hardin1968tragedy} in economics and the social sciences.

In light of these growing challenges, this paper focuses on (i) characterizing the 
impact of self-interested decision making on system performance and (ii) deriving 
locally implementable mechanisms to help mitigate these inefficiencies.  
Before delving into our specific contributions, we begin with two motivating examples: 
incentive design in congestion games and distributed coordination of multiagent systems.
%
Both these problem settings have been widely studied in the operations research and game theory literature as they capture the adverse effects of local decision making on system wide performance. 
Applications include routing in traffic and communication networks~\citep{colini2020selfish,scarsini2018dynamic}, 
distributed resource allocation~\citep{gkatzelis2016optimal,harks2011worst}, 
credit assignment in teams~\citep{kleinberg2011mechanisms}, among many others.  
The two problems detailed below apply to vastly different fields of research and yet are intimately related as 
they prompt the same set of questions: how can we characterize and optimize the system performance 
as measured by the \emph{price of anarchy}?

\subsection{Motivating Example \#1 : Incentive design in congestion games} \label{sec:motivating_example}
A widely studied model for self-interested resource allocation problems is that of (atomic) congestion games~\cite{rosenthal1973class}.
A congestion game consists of a set of users $N=\{1, \dots, n\}$ sharing the use 
of a common set of resources $\ree$, where each resource $r \in \ree$ is 
associated with a congestion function $c_r:\{1, \dots, n\} \rightarrow \arr$.  
The term $c_r(k)$ identifies the cost a user experiences for selecting resource $r$ given that there are $1 \leq k \leq n$ users concurrently selecting resource $r$. 
Further, each user $i \in N$ is associated with a given action set 
$\aee_i \subseteq 2^{\ree}$ that meets the individual needs. 
Within the context of traffic routing, for example, an action $a_i \in \aee_i$ 
describes a path in the network connecting the user's source to destination.  
Given an admissible allocation of users to resources 
$a = (a_1, a_2, \dots, a_n) \in \aee = \aee_1 \times \dots \times \aee_n$, 
the system cost describes the sum of the costs incurred by all users, i.e.,
\begin{equation} \label{eq:systemcost}
    C(a) = \sum_{i\in N}\sum_{r\in a_i} c_r(|a|_r),
\end{equation} 
where $|a|_r = |\{i \in N: r \in a_i\}|$ denotes the number of users selecting resource 
$r$ in allocation $a$. 
In this setting, an optimal allocation of users to resources consists in 
$a^{\rm opt} \in \arg \min_{a\in\cal{A}} \ C(a).$

One of the fundamental challenges associated with the allocation of resources in this problem setting is that users are often modeled as selfish decision makers. 
Specifically, each user $i \in N$ independently chooses an action $a_i \in \aee_i$ with the aim of minimizing the individual cost $J_i : \aee \rightarrow \arr$ incurred over the selected resources, i.e.,
\begin{equation}
    J_i(a) = \sum_{r \in \ree} c_r(|a|_r).
\end{equation} 
The resulting allocation is then suitably described as a pure Nash equilibrium $\nash{a}\in\cal{A}$ of the game, or a generalization thereof (e.g., mixed Nash, correlated/coarse correlated equilibria). 
Accordingly, there has been significant research seeking to quantify the quality of Nash equilibria relative to the optimal allocation. 
This is typically measured using the notion of \emph{price of anarchy}~\cite{koutsoupias1999worst}, {i.e., the worst case ratio $C(\nash{a}) / C(\opt{a})$ across a family of games.}

The analysis of the price of anarchy in congestion games has a rich history and clearly 
demonstrates the inefficiencies associated with self interested decision making
\cite{aland2011exact,awerbuch2005price,christodoulou2005price,suri2007selfish}. 
Given these inefficiencies, there is also significant research interest in the design 
of incentives that alter the users' experienced costs, thereby influencing the set of 
Nash equilibria and, thus, improving the price of anarchy 
\cite{bilo2016dynamic,caragiannis2010taxes,christodoulou2004coordination,gkatzelis2016optimal,kleer2019tight,paccagnan2019incentivizing}.
Within this context, each resource $r \in \ree$ is commonly associated with an incentive function $\tau_r : \{1, \dots, n\} \rightarrow \arr$, where $\tau_r(k)$ denotes the incentive imposed on resource $r$ when there are $k$ users selecting it.\footnote{We will use the terminology of ``tax'' and ``rebate'' when referring to incentives satisfying $\tau_r(k)\geq 0$ and $\tau_r(k)\leq 0$ for all $r$ and $k$, respectively.}
As a result, user $i \in N$ experiences a cost accounting for both the congestion on the resources and the imposed incentives, i.e., 
\begin{equation} \label{eq:33}
    J_i(a) = \sum_{r \in a_i} \big[ c_r(|a|_r) + \tau_r(|a|_r) \big].
\end{equation}
It should be stressed that, while the incentives modify the users' cost functions, they do not alter the assessment of the system cost, which still takes the form in~\eqref{eq:systemcost}. Thus, unless all incentive are identically zero, the sum of the users' costs in~\eqref{eq:33} does not equal the system cost in \eqref{eq:systemcost}. %

The objective of a system operator is to design admissible incentives, i.e., functions 
$\{\tau_r\}_{r\in\cal{R}}$ that satisfy a viable constraint on monetary budget, to 
improve the quality of the resulting collective behaviour. 
While this objective may seem deceptively straightforward, it requires the following 
theoretical advances which are currently unresolved:

\vspace{0.2cm}\noindent \textbf{{-- Characterization}:} 
What is the price of anarchy for a given set of incentives $\{\tau_r\}_{r \in \ree}$?

\vspace{0.2cm}\noindent \textbf{{-- Optimization}:}   
What are admissible incentives $\{\tau_r\}_{r \in \ree}$ that optimize the price of anarchy?

\subsection{Motivating example \#2 : Distributed coordination of multiagent systems}
\label{sec:motivatingexample2}
Alternatively, in systems where the set of users is a group of autonomous, 
computer controlled entities, the users' decision making processes can be 
explicitly selected by the system designer. 
Resource allocation problems represent a particular class of such multiagent 
systems. 
In a resource allocation problem a set of $n$ users, $N=\{1,\dots,n\}$, must be 
allocated to a set of resources $\cal{R}$. 
Each user $i\in N$ has a corresponding set of permissible actions, $\cal{A}_i$. 
For a given allocation $a=(a_1,\dots,a_n)\in\cal{A}_1 \times \dots \times \cal{A}_n=\cal{A}$, we consider a system welfare with the following separable form:
\[ W(a) = \sum_{r\in\cal{R}} W_r(|a|_r), \]
where we refer to $W_r:\{1,\dots,n\}\to \bb{R}$ as the resource welfare function on resource $r$, and denote with $\opt{a} \in \underset{a\in\cal{A}}{\arg \max} \ W(a)$ an optimal allocation. 
Applications of this model are not limited to distributed sensing \cite{gairing2009covering,zhu2013distributed}, 
resource allocation \cite{dai2015game,gkatzelis2016optimal} 
and task assignment \cite{augustine2015dynamics,chopra2017distributed}.

Requirements for scalability and security of multiagent systems compounded with 
constraints on the users' communication and computation capabilities make 
centralized coordination undesirable or even impossible.
Thus, there has recently been increased interest in the distributed coordination 
of such systems \cite{chandan2021tractable,gairing2009covering,gkatzelis2016optimal,harks2011worst}, 
where the users determine their actions according to a prescribed local decision making process.
A natural paradigm for the design of distributed coordination algorithms consists in 
i) the assignment of a local utility function to each user, and ii) the choice of a 
learning rule that specifies each user's decision making process in light of the perceived utility 
\cite{marden2013distributed,shamma2007cooperative}.
The class of distributed welfare games~\cite{marden2013distributed} provides a framework 
for studying resource allocation problems under this lens. 
In this context, each user $i\in N$ is associated with a utility function $U_i:\cal{A}\to\bb{R}$
of the following form:
\[ U_i(a) = \sum_{r\in a_i} F_r(|a|_r), \]
where the utility generating functions $F_r:\{1,\dots,n\}\to \bb{R}$, $r\in\cal{R}$, are \emph{subject to our design}. As such, the sum of the users' utilities need not be equal to the system welfare.
Remarkably, the performance (formally, the approximation ratio) of many learning rules including best response and no-regret dynamics matches the price of anarchy of a corresponding game whereby each user in $N$ is associated to the action set $\mathcal{A}_i$ and utility ${U}_i$, see \cite{roughgarden2015intrinsic}.
Thus, in order to derive efficient coordination algorithms, we must address the following questions: 

\vspace{.2cm}\noindent {\bf{-- Characterization}:} 
What is the price of anarchy of a given set of utility generating functions?

\vspace{0.2cm}\noindent \textbf{{-- Optimization}:}   
What are utility generating functions that optimize the price of anarchy?

\subsection{Our contributions} \label{sec:contributions}
The focus of this manuscript is on developing an exact, computationally efficient technique to address both the \emph{characterization} and \emph{optimization} questions highlighted above for a broad class of games that includes congestion games and distributed welfare games. The specific contributions associated with this paper are as follows:
\begin{enumerate}[leftmargin=*]
    \item In Section~\ref{sec:cost_minimization_games}, we propose a generalization of the smoothness framework introduced by Roughgarden~\cite{roughgarden2015intrinsic}. We show that for \emph{any} cost minimization game (or analagous welfare maximization game), this new framework gives an improved bound on the price of anarchy when compared to the original smoothness framework (\cref{thm:generalized_smoothness}).
    \item Our second result focuses on a generalization of congestion games where the system-level objective is not necessarily equal to the sum of the users' individual welfares. Here, we demonstrate that our framework \emph{tightly} characterizes the price of anarchy for any such game (Theorem~\ref{thm:gs_gcg}). This is in contrast to the original smoothness framework which provides only an upper bound on the price of anarchy in these settings. 
    \item Our third result shifts attention to the efficient characterization of the price of anarchy. In particular, the problem of characterizing the price of anarchy is transformed to the problem of computing an optimal set of parameters $(\lambda,\mu)$ over a given admissible space. In Theorem~\ref{thm:gs_gcg}, we provide a tractable linear program that can characterize the exact price of anarchy for any set of generalized congestion games using our framework. 
    \item Our fourth result focuses on optimizing the price of anarchy. Specifically, we show that the linear program for computing the price of anarchy can be modified to derive incentives that minimize the price of anarchy in generalized congestion games (\cref{thm:optimize_poa}). 
    \item Lastly, in \cref{sec:illustrative_examples}, we demonstrate the applicability of our methodology to the problems of incentive design in congestion games and utility design in distributed welfare games introduced in \cref{sec:motivating_example,sec:motivatingexample2}. Additionally, we show how our approach recovers and unifies a variety of existing results on the price of anarchy.
\end{enumerate}
\noindent
All above results extend unchanged to mixed Nash, correlated and coarse correlated equilibria.

\subsection{Related literature}

The notion of price of anarchy was introduced by Koutsoupias and Papadimitriou in 1999 as a metric to quantify equilibrium performance~\cite{koutsoupias1999worst}.
While a number of works have initially derived bounds on such metric, the breakthrough in the analysis of the price of anarchy came with the introduction of smoothness style arguments in two studies on atomic congestion games with affine latency functions \cite{awerbuch2005price,christodoulou2005price}.
The smoothness framework was later formalized and generalized by Roughgarden~\cite{roughgarden2015intrinsic}. 
This approach has not only proven to be useful for characterizing the efficiency of many classes of equilibria \citep{aland2011exact,caragiannis2015bounding,nisan2007algorithmic}, it has also been applied more broadly in problems including learning \citep{foster2016learning} and mechanism design \citep{syrgkanis2013composable}.
The smoothness framework provides several advantages when deriving bounds on the price of anarchy: it is tight for well studied families of games; and, it consists of a standard set of linear inequalities that govern the price of anarchy bound.
However, as we show in this manuscript, the original smoothness argument does not provide exact bounds on the price of anarchy in settings when the social welfare is not aligned with the system-level objective, i.e., $\sum_i J_i(a) \neq C(a)$, and thus provides an inaccurate approach to the problems of incentive design and utility design. 
The generalization of smoothness proposed in this manuscript resolves these deficiencies, \mbox{while retaining the strengths of the original smoothness approach.}

The notion of generalized smoothness presented in this work is most similar to the style of argument used by Gairing~\cite{gairing2009covering} to quantify the price of anarchy of covering problems. 
This work also builds upon the results of Paccagnan et al.~\cite{paccagnan2018distributed}, who provide a linear programming framework for characterizing and optimizing the efficiency of pure Nash equilibria in restricted classes of resource allocation games. 
Generalized smoothness permits a non-trivial extension of their framework: we are now able to construct linear programs for computing and optimizing the coarse-correlated equilibrium efficiency, relative to a broader class of problems that includes the class of atomic congestion games.
For an in-depth study on optimal local incentive design within the class of atomic congestion games, we refer the interested reader to Paccagnan et al.~\cite{paccagnan2019incentivizing}.

Our derivation of a linear programming technique to compute upper and lower bounds on the price of anarchy is inspired by the \emph{primal-dual approach}~\cite{bilo2012unifying,nadav2010limits}.
The primal-dual approach was used in Nadav and Roughgarden~\cite{nadav2010limits} to understand when the smoothness bound from~\cite{roughgarden2015intrinsic} is exact.
It was then specialized to the class of weighted congestion games in~\cite{bilo2012unifying}, applied to bound the efficiency of approximate Nash equilibria in~\cite{bilo2016robustness} and used to identify the best achievable price of anarchy for weighted polynomial congestion games with incentives based on the optimal allocation in~\cite{bilo2016dynamic}.
It is important to note that these prior works also propose linear programming techniques for computing upper and lower bounds on the price of anarchy.
However, as we discuss in \cref{sec:comparison_with}, our technique provides an \emph{exact} and \emph{computationally tractable} characterization of the price of anarchy, while the techniques introduced in the above works are either inexact (i.e., the bounds do not always match) or are not computationally efficient (i.e., the complexity of computing the bounds grows exponentially in the number of users $n$) for the class of generalized congestion games considered in this paper.
The improvements we obtain stem from the formalization of the generalized smoothness framework in addition to the use of a succinct parameterization that guarantees tightness of the price of anarchy bound.
As the price of anarchy bound we obtain is exact, there is no further analysis required: our linear program automatically generates a worst case game instance (lower bound) and a matching generalized smoothness argument (upper bound).
Furthermore, our framework can be modified to efficiently compute incentives and utilities that optimize the price of anarchy.


\subsection{Outline}
This article is organized as follows.
\cref{sec:cost_minimization_games} defines the class of games and the performance 
metrics that we consider throughout this paper, reviews the original notion of 
smoothness \cite{roughgarden2015intrinsic} and defines the novel generalized 
smoothness argument.
\cref{sec:slcsg} refines our study to the class of generalized congestion games,
presents our results relating to the characterization of tight and tractable bounds 
on the price of anarchy using the primal-dual approach in conjunction with a 
novel game parameterization and the derivation of optimal incentives under this 
specialized game model.
\cref{sec:welfare_maximization_games} presents analogous results for the 
welfare maximization problem setting without proof.
\cref{sec:illustrative_examples} applies our theoretical results to the problems of 
incentive design in congestion games and utility design in distributed welfare games.
\cref{sec:conclusion} includes our conclusions and a brief discussion on future work.


\section{Generalized Smoothness in Cost Minimization Games} \label{sec:cost_minimization_games}
This section introduces the class of games and performance metrics used throughout this paper.  
We proceed to review the smoothness framework from 
Roughgaren~\cite{roughgarden2015intrinsic} and highlight its limitations.  
We then introduce a revised framework, termed generalized smoothness, that 
alleviates these limitations and improves upon the efficiency guarantees 
provided by the original smoothness framework.


\subsection{Cost minimization games} \label{sec:model}
We consider the class of cost minimization problems in which there is a set of users 
$N = \{1, \dots, n\}$, and where each user $i \in N$ is associated with a given action set 
$\aset_i$ and a cost function $J_i:\aset \to \mathbb{R}$. The system cost induced by an allocation 
$a = (a_1, \dots, a_n) \in \aset = \aset_1 \times \dots \times \aset_n$ is measured by the function 
$C:\aset \to \mathbb{R}_{>0}$, and an optimal allocation satisfies
\begin{equation} \label{eq:optimization-problem}
    \opt{a} \in \argmin_{a \in \aset} C(a).
\end{equation}
We represent a cost minimization game as defined above as a tuple 
$G = (N, \aset, C, \mathcal{J})$, where $\mathcal{J} = \{J_1, \dots, J_n \}$.  
Note that the example highlighted in \cref{sec:motivating_example} represents 
a special class of cost minimization games, where the users' cost functions 
and the system cost are separable over a given set of shared resources.

The main focus of this work is on characterizing the degradation in system wide performance 
resulting from local decision making. 
To that end, we focus on the solution concept of (pure) 
Nash equilibrium as a model of the emergent behaviour in such systems. 
A Nash equilibrium is defined as any allocation $\nash{a} \in \aset$ such that
\begin{equation} \label{eq:equilibrium_conditions}
    J_i(\nash{a}) \leq J_i(a_i, \nash{a}_{-i}) \quad \forall a_i \in \aset_i, \forall i \in N.
\end{equation}
For a given game $G$, let $\mathrm{NE}(G)$ denote the set of all allocations $a \in \aset$ 
that satisfy Equation~\eqref{eq:equilibrium_conditions}. 
Assuming the set $\mathrm{NE}(G)$ is non-empty, 
we define the \textit{price of anarchy} of the game $G$ as
\begin{equation} \label{eq:poa}
    \poa(G) := \frac{\max_{a \in \mathrm{NE}(G)} C(a)}{\min_{a \in \aset} C(a)} \geq 1.
\end{equation}
The price of anarchy represents the ratio between the costs of the worst-performing 
pure Nash equilibrium in the game $G$, and the optimal allocation.
For a given class of cost minimization games $\cg$, which may contain infinitely many game instances, 
we further define the price of anarchy as,
\begin{equation} \label{eq:poa_class}
    \poa(\cg) := \sup_{G \in \cg} \poa(G) \geq 1.
\end{equation}
Note that a lower price of anarchy corresponds to an improvement in worst case equilibrium performance
and $\poa(\cg) = 1$ implies that all Nash equilibria of all games $G \in \cg$ are optimal.

\subsection{Smoothness in cost minimization games} \label{sec:smoothness_def}
The framework of $(\lambda,\mu)$-smoothness, introduced in \cite{roughgarden2015intrinsic}, 
is widely used in the existing literature aimed at characterizing the price of anarchy 
over various classes of games. 
A cost minimization game $G$ is termed ($\lambda, \mu$)-smooth if the following two conditions are met:
\begin{enumerate}
    \item[(i)] For all $a \in \aset$, we have $\sum_{i=1}^n J_i(a) \geq C(a)$;
    \item[(ii)] For all $a,a' \in \aset$, there exist $\lambda>0$ and $\mu<1$ such that
    \begin{equation}\label{eq:traditional_smoothness}
        \sum_{i \in N} J_i(a_i', a_{-i}) \leq \lambda C(a') + \mu C(a).
    \end{equation}
\end{enumerate}
If a game $G$ is ($\lambda, \mu$)-smooth, then the price of anarchy of game $G$ is upper bounded by
\[ \poa(G) \leq \frac{\lambda}{1-\mu}. \]
Observe that if all the games in a class $\cg$ are shown to be ($\lambda,\mu$)-smooth, 
then the price of anarchy of the class $\poa(\cg)$ is also upper bounded by $\lambda/(1-\mu)$.
We refer to the best upper bound obtainable using a smoothness argument on a 
given class of games $\cg$ as the \textit{robust price of anarchy}, i.e.,
\begin{equation} \label{eq:rpoa}
    \rpoa(\cg) := \inf_{\lambda > 0, \mu < 1} \left\{ \frac{\lambda}{1-\mu} \text{ s.t. Equation~\eqref{eq:traditional_smoothness}  holds } \forall G \in \cg \right\}.
\end{equation}
It is important to note that the robust price of anarchy represents only an upper bound 
on the price of anarchy, i.e., for any class of ($\lambda,\mu$)-smooth games $\cg$, it holds
that $\poa(\cg) \leq \rpoa(\cg)$, where it could be that $\poa(\cg) <\rpoa(\cg)$.


\subsection{Generalized smoothness in cost minimization games} \label{sec:cmg}
In this section, we provide a generalization of the smoothness framework,
termed \emph{generalized smoothness}.  
We will then proceed to show how this new framework provides tighter efficiency bounds 
and covers a broader spectrum of problem settings than the original smoothness framework,
defined in the previous section.

\begin{definition}[Generalized smoothness] \label{def:generalized_smoothness}
    The cost minimization game $G$ is ($\lambda,\mu$)-generalized smooth if, for any two 
    allocations $a, a' \in \aset$, there exist $\lambda >0$ and $\mu < 1$ satisfying,
    \begin{equation} \label{eq:generalized_smoothness}
        \sum^n_{i=1} J_i(a_i', a_{-i}) - \sum^n_{i=1} J_i(a) + C(a) 
                \leq \lambda C(a') + \mu C(a).
    \end{equation}
\end{definition}

Note that we maintain the notation of ($\lambda,\mu$) as in the original notion of 
smoothness for ease of comparison.
In the specific case when $\sum^n_{i=1} J_i(a) = C(a)$ for all $a \in \aset$, observe that 
the smoothness conditions in Equation~\eqref{eq:generalized_smoothness} are equivalent 
to the original smoothness conditions in Equation~\eqref{eq:traditional_smoothness}.
As with Equation~\eqref{eq:rpoa}, we define the \emph{generalized price of anarchy} of a 
class of cost minimization games $\cg$ as the best upper bound obtainable using a 
generalized smoothness argument, i.e.,
\begin{equation} \label{eq:gpoa}
    \gpoa(\cg) := \inf_{\lambda > 0, \mu < 1} \left\{ \frac{\lambda}{1-\mu} 
        \text{ s.t. Equation~\eqref{eq:generalized_smoothness}  holds } \forall G \in \cg \right\}.
\end{equation}

In our first result we show that (i)~price of anarchy bounds under the generalized smoothness framework
follow in the same way as the original smoothness framework without the restriction that 
$\sum_{i=1}^{n}J_i(a) \geq C(a)$ for all $a \in \aee$ and (ii)~the generalized smoothness framework 
provides stronger bounds on the price of anarchy than the original smoothness framework
whenever both are defined.

\begin{proposition} \label{thm:generalized_smoothness}
For any ($\lambda$,$\mu$)-generalized smooth game $G$,
the following statements hold:

\begin{enumerate}[label={(\roman*)},leftmargin=*]
    \item {The price of anarchy of $G$ is upper bounded as $\poa(G)\leq\lambda/(1-\mu)$.}
    %
    \item {If the game $G$ is ($\lambda$,$\mu$)-smooth, then $\rpoa(G)\geq\gpoa(G)\geq\poa(G)$.
    Furthermore, if $\sum_{i=1}^n J_i(a) > C(a)$ holds for all $a \in \aset$, then $\rpoa(G)>\gpoa(G)\geq\poa(G)$.}
\end{enumerate}
\end{proposition}

\proof{Proof.}
For the proof of statement~(i), observe that, for all $\nash{a} \in \mathrm{NE}(G)$ and $\opt{a} \in \mathcal{A}$,
\begin{equation}\label{eq:gsmoothinequality}
    \begin{split}
        C(\nash{a}) \leq \sum_{i=1}^n J_i(\opt{a}_i, \nash{a}_{-i}) - \sum_{i=1}^n J_i(\nash{a}) + C(\nash{a}) 
                \leq {\lambda}\,C(\opt{a}) + {\mu}\,C(\nash{a}).
    \end{split}
\end{equation}
The inequalities hold by Equations~\eqref{eq:equilibrium_conditions} and~\eqref{eq:generalized_smoothness}, respectively. 
Rearranging gives the result.

The remainder of the proof focuses on statement~(ii).
Since the condition $\sum^n_{i=1} J_i(a) \geq C(a)$ for all $a \in \aset$ implies that any pair 
of ($\lambda,\mu$) satisfying Equation~\eqref{eq:traditional_smoothness} necessarily satisfies 
Equation~\eqref{eq:generalized_smoothness}, we note that the generalized price of anarchy is less than
or equal to the robust price of anarchy, i.e., $\rpoa(G) \geq \gpoa(G) \geq \poa(G)$.

Note that for any game $G = (N,\aee,\cee,\jee)$ with $\sum^n_{i=1} J_i(a) > C(a)$ for all 
$a \in \aset$ there must exist a uniform scaling factor $0 < \gamma < 1$ such that 
$\sum_{i=1}^n \gamma J_i(a) \geq C(a)$, but for which the price of anarchy remains the same, i.e., 
for $G' = (N,\aee,\cee,\jee')$ where $\jee' = \{\gamma J_1, \dots, \gamma J_n\}$, it holds that 
$\poa(G') = \poa(G)$.
The price of anarchy remains the same despite the rescaling, 
because the inequalities in Equation~\eqref{eq:equilibrium_conditions} are unaffected by a positive scaling factor 
(i.e., $\mathrm{NE}(G) = \mathrm{NE}(G')$),
and because the optimal cost remains unchanged since the scaling does not impact the system cost. 
Further, one can verify from Equation~\eqref{eq:traditional_smoothness} that $\rpoa(G) > \rpoa(G')$, 
and thus 
$ \rpoa(G) > \rpoa(G') \geq \poa(G') = \poa(G)$.
Finally, we know that $\gpoa(G')$ is less than or equal to $\rpoa(G')$ and can 
verify from Equation~\eqref{eq:generalized_smoothness} that $\gpoa(G) = \gpoa(G')$.
Thus, 
\[ \rpoa(G) > \rpoa(G') \geq \gpoa(G') = \gpoa(G) \geq \poa(G). \]

\endproof

Further comparisons between the original notion of smoothness and generalized 
smoothness can be made, as summarized by the following observations.
These observations are stated without proof for brevity, but can easily be verified by the reader.  

\vspace{.2cm}

\noindent \emph{-- Observation \#1}: The price of anarchy and generalized price of anarchy 
are shift-, and scale-invariant, i.e., for any given $\gamma > 0$ and 
$(\delta_1, \dots, \delta_n) \in \mathbb{R}^n$,
\begin{align*}
    \poa( (N, \aset, C, \{J_i\}_{i=1}^n) ) 
            &= \poa( (N, \aset, C, \{ \gamma J_i + \delta_i \}_{i=1}^n) ), \\
    \gpoa( (N, \aset, C, \{J_i\}_{i=1}^n) ) 
            &= \gpoa( (N, \aset, C, \{ \gamma J_i + \delta_i \}_{i=1}^n) ).
\end{align*}
Neither of these properties hold for the robust price of anarchy.

\vspace{.2cm}

\noindent \emph{-- Observation \#2}: The robust price of anarchy is optimized 
by budget-balanced user cost functions, i.e., $\sum_{i\in N} J_i(a) = C(a)$ for all $a \in \aset$.   
In general, this does not hold for the price of anarchy and generalized price of anarchy.

\vspace{.2cm}

\noindent \emph{-- Observation \#3}: 
For a given cost minimization game $G$, we define an average coarse-correlated equilibrium as a 
probability distribution $\sigma \in \Delta(\aset)$ satisfying, for all $a' \in \aset$,
\begin{equation} \label{eq:acce}
    \mathbb{E}_{a \sim \sigma} \left[ \sum_{i=1}^N J_i(a) \right] 
        := \sum_{a \in \aset} \left[  \sigma_a \sum_{i=1}^N J_i(a) \right] 
                \leq \sum_{a \in \aset} \left[ \sigma_a \sum_{i=1}^N J_i(a_i, a'_{-i}) \right],
\end{equation}
where $\sigma_a \in [0,1]$ is the probability associated with action $a \in \aset$ in the 
distribution $\sigma$.
Note that the set of average coarse correlated equilibria contains all of the game's 
pure Nash equilibria, mixed Nash equilibria, correlated equilibria and coarse-correlated 
equilibria~\cite{roughgarden2015intrinsic}. 
The generalized price of anarchy tightly characterizes the average 
coarse correlated equilibrium performance of any cost minimization game $G$, 
and, thus, of any class of cost minimization games~$\cg$.
The proof follows identically to the result by Nadav and Roughgarden~\cite{nadav2010limits} 
that proves this claim for the robust price of anarchy under an alternative definition of 
average coarse correlated equilibrium. 
The two equilibrium definitions match for games with $\sum_{i=1}^n J_i(a) = C(a)$.


\section{Generalized Smoothness in Generalized Congestion Games} \label{sec:slcsg}%

The previous section introduced the framework of generalized smoothness and showed 
that the resulting generalized price of anarchy provides improved bounds on the 
price of anarchy when compared with the robust price of anarchy.  
However, deriving the generalized price of anarchy still requires solving for the 
optimal $\lambda$ and $\mu$ given in Equation~\eqref{eq:gpoa}.
In this section, we show that the optimal parameters $\lambda$ and $\mu$ for a 
generalization of the well-studied class of congestion games can be computed as 
solutions of a tractable linear program.
Furthermore, we demonstrate that the generalized price of anarchy tightly 
characterizes the price of anarchy for this important class of games,
extending the canonical results on the robustness of the price of anarchy from 
Roughgarden~\cite{roughgarden2015intrinsic} to the broader class of generalized 
congestion games that we present below.

\subsection{Generalized congestion games}%
In this section, we consider a generalization of the congestion game framework that consists of a 
user set $N = \{1, \dots, n\}$ and a resource set $\resset$. 
Given an allocation $a \in \aset$, the system cost and user cost functions have the following 
separable structure:
\begin{align}
    C(a) &= \sum_{r \in \resset} C_r(|a|_r),         \label{eq:gc_system_cost} \\
    J_i(a_i, a_{-i}) &= \sum_{r \in a_i} F_r(|a|_r), \label{eq:gc_user_cost}
\end{align}
where $C_r : \{0, 1, \dots, n\} \rightarrow \arr_{\geq 0}$ and 
$F_r : \{1, \dots, n\} \rightarrow \arr$ define the resource cost functions and cost generating 
functions, respectively. 
We will denote a congestion game by the tuple $G = (N, \ree, \aee, \{C_r, F_r\}_{r \in \ree} )$.
This game model covers many of the existing models studied in the game 
theoretic literature, including congestion games~\citep{rosenthal1973class}.

\begin{example}[Congestion Games] \label{ex:congestion_games}
In congestion games, each resource $r \in \ree$ is associated with a 
congestion function $c_r : \{1, \dots, n\} \rightarrow \arr_{\geq 0}$.  
Here, the resource cost and cost generating functions are
$C_r(k) = k \cdot c_r(k)$ and $F_r(k) = c_r(k)$ for any $k \geq 1$. 
Note that $C_r(k) = k \cdot F_r(k)$ for this case, hence the definitions of 
smoothness and generalized smoothness coincide.
\end{example}

\begin{example}[Congestion Games with Incentives] \label{ex:congestion_games_with_tolls}
When incentives are introduced into the congestion game setup, 
each resource $r \in \ree$ is also associated with an incentive 
function $\tau_r : \{1, \dots, n\} \rightarrow \arr$. 
For this class of games, the resource cost and cost generating 
functions take on the form where $C_r(k) = k \cdot c_r(k)$ and 
$F_r(k) = c_r(k) + \tau_r(k)$ for any $k \geq 1$. 
Note that for the case when $\tau_r(k)>0$ for all $r$ (i.e., taxes), 
then $C_r(k)<k\cdot F_r(k)$ and the generalized price of anarchy provides a 
strictly closer bound on the price of anarchy than the robust price of anarchy, by 
\cref{thm:generalized_smoothness}. 
Furthermore, when $\tau_r(k)<0$ for all $r$ (i.e., rebates),
then $C_r(k)>k\cdot F_r(k)$ and the original smoothness framework is inadmissible, 
by definition. 
\end{example}

\subsection{Tight price of anarchy for generalized congestion games}
Our goal in this section is to characterize the price of anarchy for a given set of 
generalized congestion games $\gee$. We begin by defining our set of games.

\begin{definition}
A generalized congestion game $G = (N, \ree, \aee, \{C_r, F_r\}_{r \in \ree})$ is generated 
from basis function pairs $\{C^j, F^j\}$, $j=1,\dots,m$, if there 
exists a set of coefficients $\alpha^1_r, \dots \alpha^m_r \geq 0$ such that
$C_r(k) = \sum_{j=1}^m \alpha^j_r \cdot C^j(k)$ and $F_r(k) = \sum_{j=1}^m \alpha^j_r \cdot F^j(k)$ 
for any $k \in \{1, \dots, n\}$ and any $r\in \ree$.
\end{definition}

We let $\cal{G}$ denote the set of all generalized congestion games with a maximum of $n$ users 
that can be generated from given basis function pairs $\{C^j, F^j\}$, $j=1,\dots,m$. 
The following example demonstrates how a limited set of basis function pairs can 
actually model a diverse set of games.

\begin{example}[Affine and Polynomial Congestion Games] \label{ex:polynomial}
One commonly studied class of congestion games is affine congestion games, 
where each resource $r \in \ree$ is associated with a cost generating 
function $F_r(k) = a_r \cdot k + b_r$ and resource cost function 
$C_r(k) = k \cdot (a_r \cdot k + b_r) = k \cdot F_r(k)$ for any $k \geq 1$, where $a_r, b_r \geq 0$. 
Observe that all admissible function pairs $\{C_r, F_r\}$ can be represented as linear combinations
of the basis function pairs 
$\{C^1, F^1\}$, $\{C^2, F^2\}$ where 
$\{C^1(k), F^1(k)\} = \{k,1\}$ (case where $a_r=0$ and $b_r=1$) and 
$\{C^2(k), F^2(k)\}=\{k^2, k\}$ (case where $a_r=1$ and $b_r=0$). 
Similarly, the function pairs $\{C_r, F_r\}$ of any polynomial congestion game of degree $d \geq 1$, 
i.e., where each resource is associated with a cost generating function of the form  
$F_r(k) = \sum_{j=1}^{d+1} \alpha^j_r \cdot k^{j-1}$ 
such that $\alpha^1_r, \dots, \alpha^{d+1}_r \geq 0$ and $C_r(k) = k \cdot F_r(k)$, 
can be represented as linear combinations of $d+1$ basis function pairs in the same 
fashion as in the affine case. 
\end{example}

The following theorem provides our main contribution pertaining to the price of anarchy 
in generalized congestion games.  
Throughout, we define $C^j(0)=F^j(0)=F^j(n+1)=0$, $j=1,\dots,m$, for ease of notation 
and without loss of generality.
Additionally, $\IR(n)$ is defined as the set of all triplets $(x,y,z) \in \{0, 1, \dots, n\}^3$ 
that satisfy: 
(i)~$1\leq x+y-z\leq n$ and $z \leq \min\{x, y\}$; 
and, (ii)~$x+y-z = n$ or $(x-z)(y-z)z = 0$.  
The structure of the set $\IR(n)$ comes from our game parameterization and will be 
fully addressed in the proof of Theorem~\ref{thm:gs_gcg}.

\begin{theorem} \label{thm:gs_gcg}
Let $\gee$ denote the set of all generalized congestion games with a maximum of $n$ users
generated from basis function pairs $\{ C^j, F^j \}$, $j=1,\dots,m$,
and let $\opt{\rho}$ be the optimal value of the following (tractable) linear program:
\begin{equation} \label{eq:characterize_poa_lp}
\begin{aligned}
    \opt{\rho} = \>
        \underset{\nu \in \mathbb{R}_{\geq 0}, \rho \in \mathbb{R}}{\text{maximize}} \quad & \rho \\
    \text{subject to:} \quad & C^j(y) - \rho C^j(x) + \nu [ (x-z) F^j(x) - (y-z) F^j(x+1) ] \geq 0, \\
    & \hspace*{125pt} \forall j=1,\dots,m, \quad \forall (x,y,z) \in \IR(n).
\end{aligned}
\end{equation}
Then, it holds that $\poa( \gee ) = \gpoa( \gee ) = 1/\opt{\rho}$.
\end{theorem}

There are two significant findings associated with Theorem~\ref{thm:gs_gcg}.  
First, observe that the generalized price of anarchy achieves a tight bound on the price of 
anarchy for any set of generalized congestion games.  
Therefore, there is no loss in characterizing the price of anarchy using the generalized 
smoothness bound.  
Second, the price of anarchy associated with a set of generalized congestion games $\gee$ 
can be characterized by means of a tractable linear program that scales linearly in its 
complexity with the number of basis function pairs, $m$, and quadratically with the number 
of users $n$.  
Thus, there are computationally efficient mechanisms for characterizing the price of anarchy 
in a given set of congestion games.

\subsection{Proof of Theorem~\ref{thm:gs_gcg}}
The following informal outline of the proof for \cref{thm:gs_gcg} is directly 
followed by the formal proof, which follows a similar structure:

\noindent\emph{-- Step 1} :
We define our game parameterization, which represents any generalized congestion game $G\in\cal{G}$ 
with $\cal{O}(mn^3)$ parameters $\theta(x,y,z,j)\geq 0$ corresponding 
with basis pairs $\{(C^j,F^j)\}$, $j=1,\dots,m$, and triplets $x,y,z\in\{0,\dots,n\}$
such that $1\leq x+y-z\leq n$ and $z\leq\min\{x,y\}$.

\noindent\emph{-- Step 2} :
For any class of generalized congestion games $\cal{G}$, we observe that an upper 
bound on the generalized price of anarchy can be computed as a fractional program 
with $m\times|\IR(n)|$ constraints under the game parameterization presented in Step 1.

\noindent\emph{-- Step 3} :
Following a change of variables, we observe that the linear program in 
Equation~\eqref{eq:characterize_poa_lp} is equivalent to the fractional program 
from Step 2.
We then provide a game $G\in\cal{G}$ with price of anarchy equal to the upper bound on 
the generalized price of anarchy, implying that ${\rm PoA}(G)\geq{\rm GPoA}(\cal{G})$.
Since ${\rm PoA}(G)\leq{\rm PoA}(\cal{G})\leq{\rm GPoA}(\cal{G})$, it must then be that 
${\rm PoA}(G)={\rm PoA}(\cal{G})={\rm GPoA}(\cal{G})$ for any class of generalized 
congestion games $\cal{G}$, concluding the proof.


\begin{proof}[Proof of~\cref{thm:gs_gcg}.]
The proof is shown in three steps, corresponding with the informal outline:

\vspace{.2cm}
\noindent\emph{-- Step 1} : 
For a given game $G\in\cal{G}$, our game parameterization is defined as follows
for allocations $a,a'\in\cal{A}$:
For every resource $r\in\cal{R}$, we define integers $x_r,y_r,z_r\geq 0$ where
$x_r=|a|_r$ is the number of users that select $r$ in $a$, 
$y_r=|a'|_r$ is the number of users that select $r$ in $a'$ and
$z_r = |\{i\in N \text{ s.t. } r\in a_i\} \cap \{i\in N \text{ s.t. } r\in a'_i\}|$
is the number of users that select $r$ in both $a$ and $a'$.
Note that $1\leq x_r+y_r-z_r\leq n$ and $z_r\leq\min\{x_r,y_r\}$ must hold for all 
$r\in\cal{R}$.
For all $x,y,z\geq 0$ such that $1\leq x+y-z\leq n$ and $z\leq\min\{x,y\}$,
and all $j=1,\dots,m$, we define the parameters
\begin{equation} \label{eq:thetas_definition}
    \theta(x,y,z,j) = \sum_{r\in\cal{R}(x,y,z)} \alpha^j_r,
\end{equation}
where $\cal{R}(x,y,z)=\{r\in\cal{R} \text{ s.t. } (x_r,y_r,z_r)=(x,y,z)\}$.
and $\alpha^j_r\geq 0$, $j=1,\dots,m$, are the coefficients in the basis representation 
of the resource cost function $C_r$ and cost generating function $F_r$.
Although the parameterization into values $\theta(x,y,z,j) \geq 0$ is of size
$\cal{O}(mn^3)$, we show in Step~2 that only $\cal{O}(mn^2)$ parameters are 
needed in the computation of the price of anarchy.

\vspace{.2cm}\noindent\emph{-- Step 2} : 
For any generalized congestion game $G\in\cal{G}$, we denote an optimal allocation 
as $\opt{a}$, and a Nash equilibrium as $\nash{a}$, i.e. $\nash{a}\in\mathrm{NE}(G)$ 
such that $\poa(G) \geq C(\nash{a})/C(\opt{a})$.
We observe that using the above definitions of $(x_r, y_r, z_r)$ for $a=\nash{a}$ and $a'=\opt{a}$, 
it follows that
\[ \sum^n_{i=1} J_i(\opt{a}_i, \nash{a}_{-i}) 
    = \sum_{r \in \resset} (y_r-z_r) F_r(x_r+1) + z_r F_r(x_r). \]
Informally, if a user $i \in N$ selects a given resource $r \in \resset$ in both $\nash{a}_i$ 
and $\opt{a}_i$, then by deviating from $\nash{a}_i$ to $\opt{a}_i$, the user does not add to the load 
on $r$, i.e., $|\opt{a}_i, \nash{a}_{-i}|_r = |\nash{a}|_r = x_r$. 
However, if $r \in \opt{a}_i$ and $r \notin \nash{a}_i$, then 
$|\opt{a}_i, \nash{a}_{-i}|_r = |\nash{a}|_r +1 = x_r+1$.

Recall that for all $r \in \cal{R}$, it must hold that $z_r \leq \min\{x_r, y_r\}$, and 
$1 \leq x_r + y_r - z_r \leq n$. 
We define the set of triplets $\iset(n) \subseteq \{0, 1, \dots, n\}^3$ as
\[ \iset(n) := \{ (x,y,z) \in \mathbb{N}^3 \text{ s.t. } 
    1 \leq x + y - z \leq n \text{ and } z \leq \min\{x, y\} \}, \]
and $\gamma(\gee)$ as the value of the following fractional program:
\begin{equation} \label{eq:ind_cons}
\begin{aligned}
    \gamma( \gee ) := \> & \inf_{\lambda > 0, \mu < 1} \quad \frac{\lambda}{1-\mu} \\
    & \text{s.t.} \quad 
        (z-x) F^j(x) + (y-z) F^j(x+1) + C^j(x) \leq \lambda C^j(y) + \mu C^j(x), \\
    & \hspace*{150pt} \forall j = 1,\dots,m, \quad \forall (x,y,z) \in \iset(n).
\end{aligned}
\end{equation}
Observe that, by Equation~\eqref{eq:thetas_definition}, the generalized smoothness 
condition in Equation~\eqref{eq:generalized_smoothness} can be rewritten within the 
context of generalized congestion games as
\begin{equation*}
\begin{aligned}
    & \sum_{(x,y,z)\in\cal{I}(n)} \sum^m_{j=1} \left[ (x-z)F^j(x)-(y-z)F^j(x+1)+C^j(x) \right] \theta(x,y,z,j) \\
    \leq \> & \sum_{(x,y,z)\in\cal{I}(n)} \sum^m_{j=1} \left[ \lambda C^j(y)+\mu C^j(x) \right] \theta(x,y,z,j)
\end{aligned}
\end{equation*}
It must then hold that for any pair $(\lambda, \mu)$ in the feasible set of the 
fractional program in Equation~\eqref{eq:ind_cons}, all games $G \in \gee$ are 
$(\lambda, \mu)$-generalized smooth, i.e., $\gamma(\gee) \geq \gpoa(\gee)$.
This is because the generalized smoothness condition for generalized congestion 
games can be expressed as a weighted sum with positive coefficients over a subset 
of the constraints in Equation~\eqref{eq:ind_cons}.

To conclude Step 2 of the proof, we show that it is sufficient to define $\gamma( \gee )$
in Equation~\eqref{eq:ind_cons} over the reduced set of constraints corresponding to 
$j \in \{1,\dots,m\}$
and triplets in $\IR(n) \subseteq \iset(n)$, where
\begin{equation*}
    \IR(n) := \{(x,y,z) \in \iset(n) \text{ s.t. } x+y-z = n \text{ or } (x-z)(y-z)z = 0\}.
\end{equation*}
For each $j \in \{1,\dots,m\}$ and any $(x,y,z) \in \iset(n)$, 
observe that the constraint in Equation~\eqref{eq:ind_cons} is equivalent to 
$y F^j(x+1) - x F^j(x) + z [ F^j(x) - F^j(x+1) ] \leq \lambda C^j(y) + (\mu - 1) C^j(x)$.
If $F^j(x+1) \geq F^j(x)$, the strictest condition on $\lambda$ and $\mu$ corresponds to the lowest 
value of $z$. Thus, $z = \max\{0, x+y-n\}$, and either $(x-z)(y-z)z = 0$ or $x+y-z = n$.
Otherwise, if $F^j(x+1) < F^j(x)$, then the largest value of $z$ is strictest, i.e.,
$z = \min\{x,y\}$ which satisfies $(x-z)(y-z)z = 0$.

\vspace{.2cm}
\noindent\emph{-- Step 3} :
In order to derive the game instances with price of anarchy matching $\gamma( \gee )$, 
it is convenient to perform the following change of variables: 
$\nu(\lambda, \mu) := 1/\lambda$ and $\rho(\lambda, \mu) := (1-\mu)/\lambda$.
For ease of notation, 
we will refer to the new variables simply as $\nu$ and $\rho$, respectively, i.e.,
$\nu = \nu(\lambda, \mu)$ and $\rho = \rho(\lambda, \mu)$.
For each $j \in \{1,\dots,m\}$ and each $(x,y,z) \in \IR(n)$, 
it is straightforward to verify that the constraints in Equation~\eqref{eq:ind_cons} 
can be rewritten in terms of $\nu$ and $\rho$ as
\[ C^j(y) - \rho C^j(x) + \nu [ (x-z) F^j(x) - (y-z) F^j(x+1) ] \geq 0. \]
Thus, the value $\gamma( \gee )$ must be equal to $1/\opt{\rho}$, 
where $\opt{\rho}$ is the value of the following linear program:
\begin{equation} \label{eq:charpoa}
\begin{aligned}
    \opt{\rho} = \> \underset{\nu \in \arr_{\geq 0}, \rho \in \arr}{\text{maximize}} \quad & \rho \\
    \text{subject to:} \quad & C^j(y) - \rho C^j(x) + \nu [ (x-z) F^j(x) - (y-z) F^j(x+1) ] \geq 0, \\
    & \hspace*{100pt} \forall j = 1,\dots,m, \quad \forall (x,y,z) \in \IR(n).
\end{aligned}
\end{equation}

It is important to note here that while $\gamma( \gee )$ is the infimum of a fractional program 
(see, e.g., Equation~\eqref{eq:ind_cons}), 
the value $\opt{\rho}$ can be computed as a maximum because the 
feasible set is bounded and closed.
Firstly, since $\gamma( \gee )$ is an upper bound on the price of anarchy,
its inverse (i.e., $\rho$) must be in the bounded and closed interval $[0,1]$. 
Additionally, one can verify that $\nu$ is not only bounded from below by 0, 
but also from above by the quantity 
\begin{equation} \label{eq:nubar}
\begin{aligned}
    \bar \nu := \> \min_{j \in \{1, \dots, m\}} \quad \underset{(x,y,z) \in \IR(n)}{\text{minimize}} \quad &
        \frac{C^j(y)}{(y-z)F^j(x+1) - (x-z)F^j(x)} \\
    \text{subject to:} \quad & (x-z) F^j(x) - (y-z) F^j(x+1) < 0 \text{ and } C^j(x) = 0,
\end{aligned}
\end{equation}
which comes from the constraints in Equation~\eqref{eq:charpoa} corresponding to triplets $(x,y,z) \in \IR(n)$ 
such that $C^j(x) = 0$ and $(x-z)F^j(x) - (y-z)F^j(x+1) < 0$.
Such a value must exist, as we assume $C^j(0) = 0$.
One can verify that any $j \in \{1,\dots,m\}$ and $(x,y,z) \in \IR(n)$ such that 
$C^j(x) = 0$ and $(x-z) F^j(x) - (y-z) F^j(x+1) \geq 0$ correspond to constraints that are 
satisfied trivially in Equation~\eqref{eq:charpoa} 
since $\nu \geq 0$, by definition, and $C^j(y) \geq 0$ for all $y = 0,1,\dots,n$, by assumption.

We denote with $\hee^j(x,y,z)$ the halfplane of $(\nu, \rho)$ values that satisfy the constraint
corresponding to $j \in \{1,\dots,m\}$ and $(x,y,z) \in \IR(n)$, i.e.,
\[ \hee^j(x,y,z) := \left\{ (\nu, \rho) \in \arr_{\geq 0} \times \arr 
    \text{ s.t. } \rho \leq \frac{C^j(y)}{C^j(x)} + \frac{1}{C^j(x)} \nu 
        \left[ (x-z) F^j(x) - (y-z) F^j(x+1) \right] \right\}. \]
The set of feasible $(\nu, \rho)$ is the intersection of these $m \times |\IR(n)|$ halfplanes.
Since the objective is to maximize $\rho$, any solution $(\opt{\nu}, \opt{\rho})$ to the 
linear program in Equation~\eqref{eq:charpoa} must be on the (upper) boundary of the feasible set.
We argue below that a solution $(\opt{\nu}, \opt{\rho})$ can only exist in one of the 
three following scenarios:
(1) at the intersection of two halfplanes' boundaries, where one halfplane has boundary line 
with positive slope, and the other has boundary line with nonpositive slope;
(2) on a halfplane boundary line with positive slope at $\opt{\nu} = \bar \nu$;
or (3) at $(\opt{\nu}, \opt{\rho}) = (0,0)$.

We denote with $\partial \hee^j(x,y,z)$ the boundary line of the halfplane $\hee^j(x,y,z)$, i.e.,
\[ \partial \hee^j(x,y,z) := \left\{ (\nu, \rho) \in \arr_{\geq 0} \times \arr 
    \text{ s.t. } \rho = \frac{C^j(y)}{C^j(x)} + \frac{1}{C^j(x)} \nu 
        \left[ (x-z) F^j(x) - (y-z) F^j(x+1) \right] \right\}. \]
Observe that the boundary lines of halfplanes corresponding to the choice $y=z=0$ have 
$\rho$-intercept equal to zero and slope $x F^j(x) / C^j(x)$.
If $F^j(x) \leq 0$ for any $j \in \{1,\dots,m\}$ and $x \in \{1, \dots, n\}$, 
then an optimal pair $(\nu, \rho)$ is trivially at the origin, i.e., 
$(\opt{\nu}, \opt{\rho}) = (0,0)$ (i.e., scenario (3) above).
Note that the $\rho$-intercept of any halfplane boundary cannot be below 0, as we only consider cost functions 
such that $C^j(k) \geq 0$ for all $k$ and all $j$.
Otherwise, the maximum value of $\rho$ occurs at the intersection of a boundary line 
with positive slope and a boundary line with nonpositive slope (i.e., scenario (1) above)
or on a boundary line with positive slope at $\nu = \bar \nu$ (i.e., scenario (2) above).
We illustrate this reasoning in~\cref{fig:optimal_scenarios}.

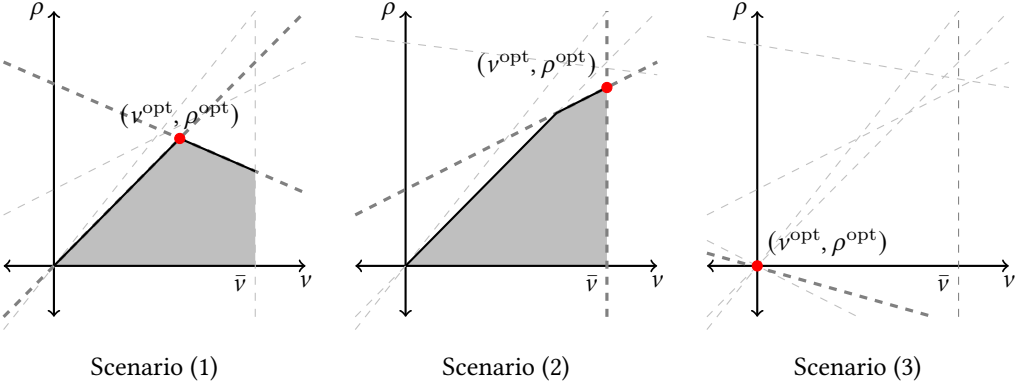
\begin{figure}
\centering
    \begin{tikzpicture}[scale=0.67]
        \fill [gray!50!white] (0,0) -- (2.5,2.5) -- (4,4-5*1.5/3.5) -- (4,0) -- cycle;
        \draw [thick, <->] (0,5) -- (0,-1);
        \draw [thick, <->] (-1,0) -- (5,0);
        \draw [thin,dashed,lightgray] (-1,1) -- (5,4);
        \draw [very thick,dashed,gray] (-1,-1) -- (5,5);
        \draw [thin,dashed,lightgray] (-1,-1.25) -- (4,5);
        \draw [very thick,dashed,gray] (-1,4) -- (5,4-6*1.5/3.5);
        \draw [thin,dashed,lightgray] (4,-1) -- (4,5);
        %
        \draw [thick] (0,0) -- (2.5,2.5) -- (4,4-5*1.5/3.5);
        \draw [red,fill] (2.5,2.5) circle [radius=0.1];
        \node at (2.5,2.5) [above] {$(\opt{\nu}, \opt{\rho})$};
        \node at (0,5) [left] {$\rho$};
        \node at (5,0) [below] {$\nu$};
        \node at (4,0) [below left] {$\bar \nu$};
        \node at (2,-2) {Scenario (1)};
    \end{tikzpicture}
    \quad
    \begin{tikzpicture}[scale=0.67]
        \fill [gray!50!white] (0,0) -- (3,3) -- (4,3.5) -- (4,0) -- cycle;
        \draw [thick, <->] (0,5) -- (0,-1);
        \draw [thick, <->] (-1,0) -- (5,0);
        \draw [very thick,dashed,gray] (-1,1) -- (5,4);
        \draw [thin,dashed,lightgray] (-1,-1) -- (5,5);
        \draw [thin,dashed,lightgray] (-1,-1.25) -- (4,5);
        \draw [thin,dashed,lightgray] (-1,4.5) -- (5,3.75);
        %
        %
        \draw [very thick,dashed,gray] (4,-1) -- (4,5);
        %
        \draw [thick] (0,0) -- (3,3) -- (4,3.5);
        \draw [red,fill] (4,3.5) circle [radius=0.1];
        \node at (4,3.5) [above left] {$(\opt{\nu}, \opt{\rho})$};
        \node at (0,5) [left] {$\rho$};
        \node at (5,0) [below] {$\nu$};
        \node at (4,0) [below left] {$\bar \nu$};
        \node at (2,-2) {Scenario (2)};
    \end{tikzpicture}
    \quad
    \begin{tikzpicture}[scale=0.67]
        \draw [thick, <->] (0,5) -- (0,-1);
        \draw [thick, <->] (-1,0) -- (5,0);
        \draw [thin,dashed,lightgray] (-1,1) -- (5,4);
        \draw [thin,dashed,lightgray] (-1,-1) -- (5,5);
        \draw [thin,dashed,lightgray] (-1,-1.25) -- (4,5);
        \draw [thin,dashed,lightgray] (-1,4.5) -- (5,3.5);
        \draw [thin,dashed,lightgray] (-1,0.5) -- (2,-1);
        \draw [very thick,dashed,gray] (-1,0.25) -- (3.5,-1);
        \draw [thin,dashed,gray] (4,-1) -- (4,5);
        %
        \draw [red,fill] (0,0) circle [radius=0.1];
        \node at (0,5) [left] {$\rho$};
        \node at (5,0) [below] {$\nu$};
        \node at (4,0) [below left] {$\bar \nu$};
        \node at (0,0) [above right] {$(\opt{\nu}, \opt{\rho})$};
        \node at (2,-2) {Scenario (3)};
    \end{tikzpicture}
\caption{\emph{The three different scenarios in which optimal solutions $(\opt{\nu}, \opt{\rho})$ 
to Equation~\eqref{eq:charpoa} can exist.}
We illustrate the reasoning behind each of the three scenarios for optimal solutions 
$(\opt{\nu}, \opt{\rho})$ to the linear program in Equation~\eqref{eq:charpoa}.
Since the objective of Equation~\eqref{eq:charpoa} is to maximize $\rho$, the optimal values will be at the (upper)
boundary of the feasible set, illustrated with a solid, bolded line in each of the examples above.
Additionally, the optimal solution $(\opt{\nu}, \opt{\rho})$ is marked by a solid, red dot in the 
illustrations above.
In Scenario~(1), on the left, $(\opt{\nu}, \opt{\rho})$ lie on the intersection of a boundary line 
with positive slope and a boundary line with nonpositive slope.
In Scenario~(2), centre, $(\opt{\nu}, \opt{\rho})$ lie on the intersection of a boundary line 
with positive slope at $\nu = \bar \nu$, which is defined in Equation~\eqref{eq:nubar}.
In Scenario~(3), on the right, there exists a halfplane boundary line with nonpositive slope and $\rho$-intercept
equal to zero, and so $(\opt{\nu}, \opt{\rho}) = (0,0)$.
Using the parameters corresponding to the halfplanes on which the pair
$(\opt{\nu}, \opt{\rho})$ lays, 
we can construct games $G \in \gee$ with $\poa(G) = 1/\opt{\rho}$ in each of 
these scenarios.}
\label{fig:optimal_scenarios}
\end{figure}

Observe that for Scenarios (1) and (2), the pair $(\opt{\nu}, \opt{\rho})$ 
is at the intersection of two boundary lines, 
which we denote as $\partial \hee^j(x,y,z)$ and $\partial \hee^{j'}(x',y',z')$.
The parameters $j, j' \in \{1,\dots,m\}$ and $(x,y,z), (x',y',z') \in \IR(n)$ satisfy the following:
\begin{equation} \label{eq:on_the_boundary}
\begin{aligned}
    \opt{\nu} [ (x-z) F^j(x) - (y-z) F^j(x+1) ] &= \opt{\rho} C^j(x) - C^j(y) , \\
    \opt{\nu} [ (x'-z') F^{j'}(x') - (y'-z') F^{j'}(x'+1) ] 
        &= \opt{\rho} C^{j'}(x') - C^{j'}(y'),
\end{aligned}
\end{equation}
because $(\opt{\nu}, \opt{\rho})$ is on both boundary lines.
Further, there exists $\eta \in [0,1]$ such that
\begin{equation} \label{eq:opposite_slopes}
    \eta \left[ (x-z) F^j(x) - (y-z) F^j(x+1) \right] 
        + (1-\eta) \left[ (x'-z') F^{j'}(x') - (y'-z') F^{j'}(x'+1) \right] = 0.
\end{equation}
Equation~\eqref{eq:opposite_slopes} holds in Scenario~(1) because one of the 
boundary lines has positive slope, i.e., $(x-z)F^j(x) - (y-z)F^j(x+1) > 0$, while 
the other has nonpositive slope, and in Scenario~(3) because one boundary line has 
positive slope while the other is the vertical line $\nu = \bar \nu$ which 
corresponds to a particular choice of $j \in \{1,\dots,m\}$ and $(x,y,z) \in \IR(n)$ 
such that $(x-z)F^j(x) - (y-z)F^j(x+1) < 0$ by Equation~\eqref{eq:nubar}.

\begin{figure}[t]
    \centering
        \begin{tikzpicture}[scale=0.75]
            \node at (0,0.75) {\Large $\resset_1$};
            \node at (0,0) {\tiny $C_r(k) = \eta C(k),$};
            \node at (0,-0.33) {\tiny $F_r(k) = \eta F(k),$};
            \node at (0,-0.67) {\tiny for all $r \in \resset_1$};
            \draw (0,0) circle [radius=2];
            \filldraw[fill=white] (0,2)         circle [radius=0.45] node {$r_1$};
            \filldraw[fill=white] (1,1.732)     circle [radius=0.45] node {$r_2$};
            \filldraw[fill=white] (1.732,1)     circle [radius=0.45] node {$r_3$};
            \filldraw[fill=white] (2,0)         circle [radius=0.45] node {$r_4$};
            \filldraw[fill=white] (1.732,-1)    circle [radius=0.45] node {$r_5$};
            \filldraw[fill=white] (1,-1.732)    circle [radius=0.45];
            \filldraw[fill=white] (0,-2)        circle [radius=0.45];
            \filldraw[fill=white] (-1,-1.732)   circle [radius=0.45];
            \filldraw[fill=white] (-1.732,-1)   circle [radius=0.45] node {$r_{n-3}$};
            \filldraw[fill=white] (-2,0)        circle [radius=0.45] node {$r_{n-2}$};
            \filldraw[fill=white] (-1.732,1)    circle [radius=0.45] node {$r_{n-1}$};
            \filldraw[fill=white] (-1,1.732)    circle [radius=0.45] node {$r_n$};
        \end{tikzpicture}
        \quad
        \begin{tikzpicture}[scale=0.75]
            \node at (0,0.75) {\Large $\resset_2$};
            \node at (0,0) {\tiny $C_r(k) = (1-\eta)C'(k),$};
            \node at (0,-0.33) {\tiny $F_r(x) = (1-\eta)F'(k),$};
            \node at (0,-0.67) {\tiny for all $r \in \resset_2$};
            \draw (0,0) circle [radius=2];
            \filldraw[fill=white] (0,2)         circle [radius=0.45] node {$r_{n+1}$};
            \filldraw[fill=white] (1,1.732)     circle [radius=0.45] node {$r_{n+2}$};
            \filldraw[fill=white] (1.732,1)     circle [radius=0.45] node {$r_{n+3}$};
            \filldraw[fill=white] (2,0)         circle [radius=0.45] node {$r_{n+4}$};
            \filldraw[fill=white] (1.732,-1)    circle [radius=0.45] node {$r_{n+5}$};
            \filldraw[fill=white] (1,-1.732)    circle [radius=0.45];
            \filldraw[fill=white] (0,-2)        circle [radius=0.45];
            \filldraw[fill=white] (-1,-1.732)   circle [radius=0.45];
            \filldraw[fill=white] (-1.732,-1)   circle [radius=0.45] node {$r_{2n-3}$};
            \filldraw[fill=white] (-2,0)        circle [radius=0.45] node {$r_{2n-2}$};
            \filldraw[fill=white] (-1.732,1)    circle [radius=0.45] node {$r_{2n-1}$};
            \filldraw[fill=white] (-1,1.732)    circle [radius=0.45] node {$r_{2n}$};
        \end{tikzpicture}
        \quad
        \begin{tikzpicture}[scale=0.67]
            \node at (-1.5,1.75) {\normalsize $i$};
            \node at (-1.5,0) {\normalsize 1};
            \node at (-1.5,-1.5) {\normalsize 2};
            \node at (-1.5,-2.75) {\normalsize $\vdots$};
            \node at (-1.5,-4) {\normalsize $n$};
            \node at (0.75,1.75) {\normalsize $\nash{a}_i$};
            \node at (0.,1) {\normalsize $\resset_1$};
            \node at (1.5,1) {\normalsize $\resset_2$};
            \draw[thick] (-0.75,2)--(-0.75,-4.5);
            \draw (0,0) circle [radius=0.5];
            \filldraw[fill=blue] (0,0.5)               circle [radius=0.075];
            \filldraw[fill=blue] (0.25,0.25*1.732)     circle [radius=0.075];
            \filldraw[fill=blue] (0.25*1.732,0.25)     circle [radius=0.075];
            \filldraw[fill=blue] (0.5,0)               circle [radius=0.075];
            \filldraw[fill=white] (0.25,-0.25*1.732)    circle [radius=0.075];
            \filldraw[fill=white] (0.25*1.732,-0.25)    circle [radius=0.075];
            \filldraw[fill=white] (0,-0.5)              circle [radius=0.075];
            \filldraw[fill=white] (-0.25,-0.25*1.732)   circle [radius=0.075];
            \filldraw[fill=white] (-0.25*1.732,-0.25)   circle [radius=0.075];
            \filldraw[fill=white] (-0.5,0)              circle [radius=0.075];
            \filldraw[fill=white] (-0.25,0.25*1.732)    circle [radius=0.075];
            \filldraw[fill=white] (-0.25*1.732,0.25)    circle [radius=0.075];
            %
            \draw (1.5+0,0) circle [radius=0.5];
            \filldraw[fill=blue] (1.5+0,0.5)               circle [radius=0.075];
            \filldraw[fill=blue] (1.5+0.25,0.25*1.732)     circle [radius=0.075];
            \filldraw[fill=blue] (1.5+0.25*1.732,0.25)     circle [radius=0.075];
            \filldraw[fill=white] (1.5+0.5,0)               circle [radius=0.075];
            \filldraw[fill=white] (1.5+0.25,-0.25*1.732)    circle [radius=0.075];
            \filldraw[fill=white] (1.5+0.25*1.732,-0.25)    circle [radius=0.075];
            \filldraw[fill=white] (1.5+0,-0.5)              circle [radius=0.075];
            \filldraw[fill=white] (1.5-0.25,-0.25*1.732)    circle [radius=0.075];
            \filldraw[fill=white] (1.5-0.25*1.732,-0.25)    circle [radius=0.075];
            \filldraw[fill=white] (1.5-0.5,0)               circle [radius=0.075];
            \filldraw[fill=white] (1.5-0.25*1.732,0.25)     circle [radius=0.075];
            \filldraw[fill=white] (1.5-0.25,0.25*1.732)     circle [radius=0.075];
            %
            \draw (0,0-1.5) circle [radius=0.5];
            \filldraw[fill=white] (0,0.5-1.5)               circle [radius=0.075];
            \filldraw[fill=blue] (0.25,0.25*1.732-1.5)     circle [radius=0.075];
            \filldraw[fill=blue] (0.25*1.732,0.25-1.5)     circle [radius=0.075];
            \filldraw[fill=blue] (0.5,0-1.5)               circle [radius=0.075];
            \filldraw[fill=blue] (0.25*1.732,-0.25-1.5)    circle [radius=0.075];
            \filldraw[fill=white] (0.25,-0.25*1.732-1.5)    circle [radius=0.075];
            \filldraw[fill=white] (0,-0.5-1.5)              circle [radius=0.075];
            \filldraw[fill=white] (-0.25,-0.25*1.732-1.5)   circle [radius=0.075];
            \filldraw[fill=white] (-0.25*1.732,-0.25-1.5)   circle [radius=0.075];
            \filldraw[fill=white] (-0.5,0-1.5)              circle [radius=0.075];
            \filldraw[fill=white] (-0.25,0.25*1.732-1.5)    circle [radius=0.075];
            \filldraw[fill=white] (-0.25*1.732,0.25-1.5)    circle [radius=0.075];
            %
            \draw (1.5+0,0-1.5) circle [radius=0.5];
            \filldraw[fill=white] (1.5+0,0.5-1.5)               circle [radius=0.075];
            \filldraw[fill=blue] (1.5+0.25,0.25*1.732-1.5)     circle [radius=0.075];
            \filldraw[fill=blue] (1.5+0.25*1.732,0.25-1.5)     circle [radius=0.075];
            \filldraw[fill=blue] (1.5+0.5,0-1.5)               circle [radius=0.075];
            \filldraw[fill=white] (1.5+0.25,-0.25*1.732-1.5)    circle [radius=0.075];
            \filldraw[fill=white] (1.5+0.25*1.732,-0.25-1.5)    circle [radius=0.075];
            \filldraw[fill=white] (1.5+0,-0.5-1.5)              circle [radius=0.075];
            \filldraw[fill=white] (1.5-0.25,-0.25*1.732-1.5)    circle [radius=0.075];
            \filldraw[fill=white] (1.5-0.25*1.732,-0.25-1.5)    circle [radius=0.075];
            \filldraw[fill=white] (1.5-0.5,0-1.5)               circle [radius=0.075];
            \filldraw[fill=white] (1.5-0.25*1.732,0.25-1.5)     circle [radius=0.075];
            \filldraw[fill=white] (1.5-0.25,0.25*1.732-1.5)     circle [radius=0.075];
            \node at (0.75,-2.75) {$\vdots$};
            %
            \draw (0,0-4) circle [radius=0.5];
            \filldraw[fill=blue] (0,0.5-4)               circle [radius=0.075];
            \filldraw[fill=blue] (0.25,0.25*1.732-4)     circle [radius=0.075];
            \filldraw[fill=blue] (0.25*1.732,0.25-4)     circle [radius=0.075];
            \filldraw[fill=white] (0.5,0-4)               circle [radius=0.075];
            \filldraw[fill=white] (0.25,-0.25*1.732-4)    circle [radius=0.075];
            \filldraw[fill=white] (0.25*1.732,-0.25-4)    circle [radius=0.075];
            \filldraw[fill=white] (0,-0.5-4)              circle [radius=0.075];
            \filldraw[fill=white] (-0.25,-0.25*1.732-4)   circle [radius=0.075];
            \filldraw[fill=white] (-0.25*1.732,-0.25-4)   circle [radius=0.075];
            \filldraw[fill=white] (-0.5,0-4)              circle [radius=0.075];
            \filldraw[fill=white] (-0.25*1.732,0.25-4)    circle [radius=0.075];
            \filldraw[fill=blue] (-0.25,0.25*1.732-4)    circle [radius=0.075];
            %
            \draw (1.5+0,0-4) circle [radius=0.5];
            \filldraw[fill=blue] (1.5+0,0.5-4)               circle [radius=0.075];
            \filldraw[fill=blue] (1.5+0.25,0.25*1.732-4)     circle [radius=0.075];
            \filldraw[fill=white] (1.5+0.25*1.732,0.25-4)     circle [radius=0.075];
            \filldraw[fill=white] (1.5+0.5,0-4)               circle [radius=0.075];
            \filldraw[fill=white] (1.5+0.25,-0.25*1.732-4)    circle [radius=0.075];
            \filldraw[fill=white] (1.5+0.25*1.732,-0.25-4)    circle [radius=0.075];
            \filldraw[fill=white] (1.5+0,-0.5-4)              circle [radius=0.075];
            \filldraw[fill=white] (1.5-0.25,-0.25*1.732-4)    circle [radius=0.075];
            \filldraw[fill=white] (1.5-0.25*1.732,-0.25-4)    circle [radius=0.075];
            \filldraw[fill=white] (1.5-0.5,0-4)               circle [radius=0.075];
            \filldraw[fill=white] (1.5-0.25*1.732,0.25-4)     circle [radius=0.075];
            \filldraw[fill=blue] (1.5-0.25,0.25*1.732-4)     circle [radius=0.075];
        \end{tikzpicture}
        \begin{tikzpicture}[scale=0.67]
            \node at (0.75,1.75) {\normalsize $\opt{a}_i$};
            \node at (0.,1) {\normalsize $\resset_1$};
            \node at (1.5,1) {\normalsize $\resset_2$};
            \draw[thick] (-0.75,2)--(-0.75,-4.5);
            \draw (0,0) circle [radius=0.5];
            \filldraw[fill=white] (0,0.5)               circle [radius=0.075];
            \filldraw[fill=white] (0.25,0.25*1.732)     circle [radius=0.075];
            \filldraw[fill=white] (0.25*1.732,0.25)     circle [radius=0.075];
            \filldraw[fill=white] (0.5,0)               circle [radius=0.075];
            \filldraw[fill=white] (0.25,-0.25*1.732)    circle [radius=0.075];
            \filldraw[fill=white] (0.25*1.732,-0.25)    circle [radius=0.075];
            \filldraw[fill=white] (0,-0.5)              circle [radius=0.075];
            \filldraw[fill=white] (-0.25,-0.25*1.732)   circle [radius=0.075];
            \filldraw[fill=white] (-0.25*1.732,-0.25)   circle [radius=0.075];
            \filldraw[fill=white] (-0.5,0)              circle [radius=0.075];
            \filldraw[fill=red] (-0.25,0.25*1.732)    circle [radius=0.075];
            \filldraw[fill=red] (-0.25*1.732,0.25)    circle [radius=0.075];
            %
            \draw (1.5+0,0) circle [radius=0.5];
            \filldraw[fill=red] (1.5+0,0.5)               circle [radius=0.075];
            \filldraw[fill=red] (1.5+0.25,0.25*1.732)     circle [radius=0.075];
            \filldraw[fill=white] (1.5+0.25*1.732,0.25)     circle [radius=0.075];
            \filldraw[fill=white] (1.5+0.5,0)               circle [radius=0.075];
            \filldraw[fill=white] (1.5+0.25,-0.25*1.732)    circle [radius=0.075];
            \filldraw[fill=white] (1.5+0.25*1.732,-0.25)    circle [radius=0.075];
            \filldraw[fill=white] (1.5+0,-0.5)              circle [radius=0.075];
            \filldraw[fill=white] (1.5-0.25,-0.25*1.732)    circle [radius=0.075];
            \filldraw[fill=white] (1.5-0.25*1.732,-0.25)    circle [radius=0.075];
            \filldraw[fill=white] (1.5-0.5,0)               circle [radius=0.075];
            \filldraw[fill=red] (1.5-0.25*1.732,0.25)     circle [radius=0.075];
            \filldraw[fill=red] (1.5-0.25,0.25*1.732)     circle [radius=0.075];
            %
            \draw (0,0-1.5) circle [radius=0.5];
            \filldraw[fill=red] (0,0.5-1.5)               circle [radius=0.075];
            \filldraw[fill=white] (0.25,0.25*1.732-1.5)     circle [radius=0.075];
            \filldraw[fill=white] (0.25*1.732,0.25-1.5)     circle [radius=0.075];
            \filldraw[fill=white] (0.5,0-1.5)               circle [radius=0.075];
            \filldraw[fill=white] (0.25*1.732,-0.25-1.5)    circle [radius=0.075];
            \filldraw[fill=white] (0.25,-0.25*1.732-1.5)    circle [radius=0.075];
            \filldraw[fill=white] (0,-0.5-1.5)              circle [radius=0.075];
            \filldraw[fill=white] (-0.25,-0.25*1.732-1.5)   circle [radius=0.075];
            \filldraw[fill=white] (-0.25*1.732,-0.25-1.5)   circle [radius=0.075];
            \filldraw[fill=white] (-0.5,0-1.5)              circle [radius=0.075];
            \filldraw[fill=white] (-0.25*1.732,0.25-1.5)    circle [radius=0.075];
            \filldraw[fill=red] (-0.25,0.25*1.732-1.5)    circle [radius=0.075];
            %
            \draw (1.5+0,0-1.5) circle [radius=0.5];
            \filldraw[fill=red] (1.5+0,0.5-1.5)               circle [radius=0.075];
            \filldraw[fill=red] (1.5+0.25,0.25*1.732-1.5)     circle [radius=0.075];
            \filldraw[fill=red] (1.5+0.25*1.732,0.25-1.5)     circle [radius=0.075];
            \filldraw[fill=white] (1.5+0.5,0-1.5)               circle [radius=0.075];
            \filldraw[fill=white] (1.5+0.25,-0.25*1.732-1.5)    circle [radius=0.075];
            \filldraw[fill=white] (1.5+0.25*1.732,-0.25-1.5)    circle [radius=0.075];
            \filldraw[fill=white] (1.5+0,-0.5-1.5)              circle [radius=0.075];
            \filldraw[fill=white] (1.5-0.25,-0.25*1.732-1.5)    circle [radius=0.075];
            \filldraw[fill=white] (1.5-0.25*1.732,-0.25-1.5)    circle [radius=0.075];
            \filldraw[fill=white] (1.5-0.5,0-1.5)               circle [radius=0.075];
            \filldraw[fill=white] (1.5-0.25*1.732,0.25-1.5)     circle [radius=0.075];
            \filldraw[fill=red] (1.5-0.25,0.25*1.732-1.5)     circle [radius=0.075];
            \node at (0.75,-2.75) {$\vdots$};
            %
            \draw (0,0-4) circle [radius=0.5];
            \filldraw[fill=white] (0,0.5-4)               circle [radius=0.075];
            \filldraw[fill=white] (0.25,0.25*1.732-4)     circle [radius=0.075];
            \filldraw[fill=white] (0.25*1.732,0.25-4)     circle [radius=0.075];
            \filldraw[fill=white] (0.5,0-4)               circle [radius=0.075];
            \filldraw[fill=white] (0.25,-0.25*1.732-4)    circle [radius=0.075];
            \filldraw[fill=white] (0.25*1.732,-0.25-4)    circle [radius=0.075];
            \filldraw[fill=white] (0,-0.5-4)              circle [radius=0.075];
            \filldraw[fill=white] (-0.25,-0.25*1.732-4)   circle [radius=0.075];
            \filldraw[fill=white] (-0.25*1.732,-0.25-4)   circle [radius=0.075];
            \filldraw[fill=red] (-0.5,0-4)              circle [radius=0.075];
            \filldraw[fill=red] (-0.25*1.732,0.25-4)    circle [radius=0.075];
            \filldraw[fill=white] (-0.25,0.25*1.732-4)    circle [radius=0.075];
            %
            \draw (1.5+0,0-4) circle [radius=0.5];
            \filldraw[fill=red] (1.5+0,0.5-4)               circle [radius=0.075];
            \filldraw[fill=white] (1.5+0.25,0.25*1.732-4)     circle [radius=0.075];
            \filldraw[fill=white] (1.5+0.25*1.732,0.25-4)     circle [radius=0.075];
            \filldraw[fill=white] (1.5+0.5,0-4)               circle [radius=0.075];
            \filldraw[fill=white] (1.5+0.25,-0.25*1.732-4)    circle [radius=0.075];
            \filldraw[fill=white] (1.5+0.25*1.732,-0.25-4)    circle [radius=0.075];
            \filldraw[fill=white] (1.5+0,-0.5-4)              circle [radius=0.075];
            \filldraw[fill=white] (1.5-0.25,-0.25*1.732-4)    circle [radius=0.075];
            \filldraw[fill=white] (1.5-0.25*1.732,-0.25-4)    circle [radius=0.075];
            \filldraw[fill=red] (1.5-0.5,0-4)               circle [radius=0.075];
            \filldraw[fill=red] (1.5-0.25*1.732,0.25-4)     circle [radius=0.075];
            \filldraw[fill=red] (1.5-0.25,0.25*1.732-4)     circle [radius=0.075];
        \end{tikzpicture}
    \caption{\emph{The game instance construction $G$ consisting of $n$ users, and two disjoint cycles $\resset_1$ and $\resset_2$, as described in the proof of~\cref{thm:gs_gcg}, Step~2
    for Scenarios~(1) and~(2).}
    Consider the set of games $\gee$, where $n$ is the maximum number of users and $\zee$ is a set of basis functions pairs, and suppose that $(\opt{\nu}, \opt{\rho})$ satisfy the conditions of Scenarios~(1) or~(2). 
    Further, suppose that the parameters for which Equation~\eqref{eq:on_the_boundary} and Equation~\eqref{eq:opposite_slopes} hold are ${C,F}, {C',F'} \in \zee$, $(x,y,z) = (4,2,0), (x',y',z') = (3,4,2) \in \IR(n)$ and $\eta \in [0,1]$.
    In the above figure, we illustrate the game $G \in \gee$ such that $\poa(G) = \poa(G') = 1/\opt{\rho}$ according to the reasoning for constructing game instances in Scenarios~(1) and~(2).
    Observe that each resource $r \in \resset_1$ has $C_r(k) = \eta C(k)$, and $F_r(k) = \eta F(k)$, whereas each resource $r \in \resset_2$ has $C_r(k) = (1-\eta) C'(k)$, and $F_r(x) = (1-\eta) F'(k)$, for all $k \in \{1, \dots, n\}$.
    Each user $i \in N$ has two actions $\nash{a}_i$ and $\opt{a}_i$, as defined in the table on the right.
    Observe that every resource in $\resset_1$ is selected by 4 users in the allocation $\nash{a} = (\nash{a}_1, \dots, \nash{a}_n)$, and 3 users in $\opt{a} = (\opt{a}_1, \dots, \opt{a}_n)$, where no user $i \in N$ has a common resource between its actions $\nash{a}_i$ and $\opt{a}_i$, i.e., $x_r = 4 = x$, $y_r = 3 = y$, and $z_r = 0 = z$ for all $r \in \resset_1$.
    Similarly, $x_r = 3 = x'$, $y_r = 4 = y'$, and $z_r = 2 = z'$, for each resource $r \in \resset_2$.
    }
    \label{fig:game_construction}
\end{figure}
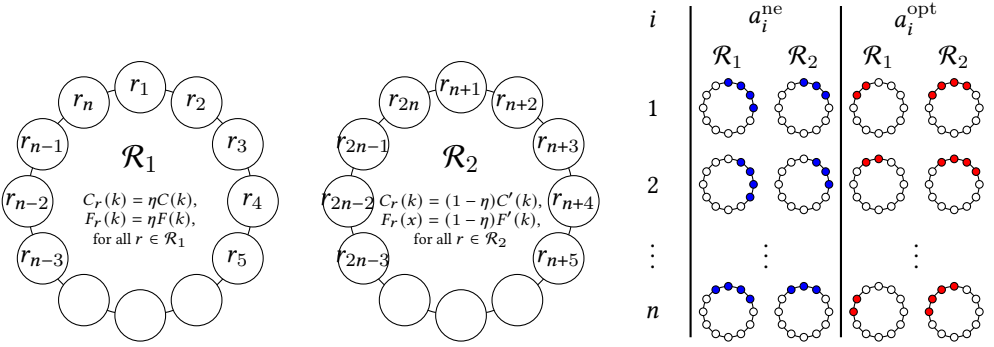

Next, for the parameters $j, j' \in \{1,\dots,m\}$, 
$(x,y,z), (x',y',z') \in \IR(n)$, and $\eta \in [0,1]$ obtained above, 
we construct a game instance $G \in \gee$ such that $\poa(G) = 1/\opt{\rho}$.
Let $\resset_1 = \{r_1, \dots, r_n\}$ and $\resset_2 = \{r_{n+1}, \dots, r_{2n}\}$ 
denote two disjoint cycles of resources. 
Every resource $r \in \resset_1$ has cost function $C_r(k) = \eta C^j(k)$, and cost generating 
function $F_r(k) = \eta F^j(k)$ for all $k$.
Meanwhile, every $r \in \resset_2$ has cost function $C_r(k) = (1-\eta) C^{j'}(k)$, and cost
generating function $F_r(k) = (1-\eta) F^{j'}(k)$ for all $k$. 
We define the user set $N = \{1, \dots, n\}$, where each user $i \in N$ has action set 
$\aset_i = \{ \nash{a}_i, \opt{a}_i \}$. 
In action $\nash{a}_i$, the user $i$ selects $x$ consecutive resources in $\resset_1$ starting 
with $r_i$, i.e. $\{r_i, r_{(i \bmod n) + 1}, \dots, r_{((i+x-2) \bmod n) + 1}\}$, and $x'$ 
consecutive resources in $\resset_2$ starting with resource $r_{n+i}$. 
In $\opt{a}_i$, user $i$ selects $y$ consecutive resources in $\resset_1$ ending with resource 
$r_{((i+z-2) \bmod n)+1}$, i.e. $\{r_{((i+z-y-1) \bmod n)+1}, \dots, r_{((i+z-2) \bmod n)+1}\}$, 
and $y'$ consecutive resources in $\resset_2$ ending with resource $r_{n+((i+z'-2) \bmod n)+1}$.
We provide an illustration of this game construction in \cref{fig:game_construction}.
Observe that $\nash{a} = (\nash{a}_1, \dots, \nash{a}_n)$ satisfies the conditions for a Nash 
equilibrium,
\begin{equation*}
\begin{aligned}
    J_i(\nash{a}) & = \eta x F^j(x) + (1-\eta) x' F^{j'}(x') \\
    & =\> \eta[ z F^j(x) + (y-z) F^j(x+1) ] + (1-\eta) [z' F^{j'}(x') + (y'-z') F^{j'}(x'+1) ] 
    = J_i(\opt{a}_i, \nash{a}_{-i})\text{,}
\end{aligned}
\end{equation*}
which holds by Equation~\eqref{eq:opposite_slopes}. 
Then, by the above equality and Equation~\eqref{eq:on_the_boundary},
\begin{align*}
    0 &= \sum_{i=1}^n J_i(\opt{a}_i, \nash{a}_{-i}) - \sum_{i=1}^n J_i(\nash{a}) \\
    &= \frac{1}{\opt{\nu}} \left[ n \cdot \eta \left[ \opt{\rho} C^j(x) - C^j(y) \right]
        + n \cdot (1-\eta) \left[ \opt{\rho} C^{j'}(x') + C^{j'}(y') \right] \right] \\
    &= \frac{1}{\opt{\nu}} \left[ \opt{\rho} C(\nash{a}) - C(\opt{a}) \right],
\end{align*}
where $\opt{a} = (\opt{a}_1, \dots, \opt{a}_n)$.
Thus, $\poa(G) = 1/\opt{\rho}$.

For Scenario~(3), observe that $\opt{\rho}=0$, and so $1/\opt{\rho}$ is unbounded.
Recall that, in this scenario, there exist $j \in \{1,\dots,m\}$ and $x \in \{1, \dots, n\}$ such that
$F^j(x) \leq 0$.
We use the basis function pair $\{C^j,F^j\}$ to construct a game $G$ with unbounded price of anarchy.
Consider a game instance with $x$ users and resource set $\ree = \{r_1, r_2\}$, 
where $x \in \{1, \dots, n\}$ is the value that minimizes the function $F(x)$, i.e., 
$F^j(x) = \min_{k \in \{1, \dots, n\}} F^j(k) \leq 0$.
Every user $i \in \{1, \dots, x\}$ has action set $\aee_i = \{ \{r_1\}, \{r_2\} \}$.
The resource $r_1$ has cost function $C_r(k) = \eta C^j(k)$ 
and cost generating function $F_r(k) = \eta F^j(k)$ for all $k$.
Similarly, the resource $r_2$ has cost function $C_r(k) = (1-\eta) C^j(k)$ 
and cost generating function $F_r(k) = (1-\eta) F(k)$.
It is straightforward to verify that, for $\eta$ approaching 0 from above, the 
allocation in which all users select $r_1$ is an equilibrium and the price of 
anarchy is unbounded.

\end{proof}

\subsection{Comparison to Existing Literature} \label{sec:comparison_with}

There has been a significant amount of research focused on characterizing the price of 
anarchy in congestion games.  
Accordingly, in this section, we position the results of Theorem~\ref{thm:gs_gcg} in the 
broader context of smoothness \cite{roughgarden2015intrinsic} and the primal-dual
approach \cite{bilo2012unifying,nadav2010limits}.  
First, it is important to recognize that both the smoothness and generalized smoothness 
frameworks can be written as linear programs for any family of games $\cal{G}$.
For example, observe that the generalized price of anarchy satisfies 
${\rm GPoA}(\cal{G})=1/\opt{\rho}$, where
\begin{equation} \label{eq:gsmooth_lp}
\begin{aligned}
    \opt{\rho} = \> 
        \underset{\nu \in \mathbb{R}_{\geq 0}, \rho \in \mathbb{R}}{\text{maximize}} \quad & \rho \\
    \text{subject to:} \quad & C(a')-\rho C(a)+\nu\left[\sum^n_{i=1} J_i(a) - \sum^n_{i=1} J_i(a'_i, a_{-i})\right] \geq 0,
        \quad \forall a,a' \in \cal{A}, \forall G\in\cal{G},
\end{aligned}
\end{equation}
which follows from Equations~\eqref{eq:generalized_smoothness} and~\eqref{eq:gpoa}
for the change of variables $\nu=1/\lambda$ and $\rho=(1-\mu)/\lambda$.
Although the price of anarchy bound obtained using the above linear program is the 
best achievable following a generalized smoothness argument, computing such a bound 
for a family of games is intractable as there may be exponentially many constraints, 
even for modest values of the maximum number of users $n$.
The novelty of the result in \cref{thm:gs_gcg} is in identifying a game 
parameterization for any set of generalized congestion games such that the number of 
linear program constraints only grows linearly in $m$ and quadratically in $n$ while 
verifying and preserving the tightness of the generalized price of anarchy bound.

Our result is inspired by several previous works, most notably Bil{\`o}~\cite{bilo2012unifying} 
and Roughgarden~\cite{roughgarden2015intrinsic}, that introduce game parameterizations to 
reduce the complexity of smoothness bounds.
We note that many of the bounds proposed in these previous works remain intractable, 
as the number of linear program constraints grows exponentially in $n$.
Nonetheless, \cite{bilo2012unifying} provides a tractable linear program 
for deriving upper bounds on the price of anarchy that has two decision variables and
$\cal{O}(n^2)$ constraints.
Here, we demonstrate that upper bounds computed using the tractable linear program in 
\cite{bilo2012unifying} are not tight, even for affine congestion games with $n=2$ users.

\begin{example}
\label{ex:bilo_loose}
For the set of affine congestion games $\cal{G}$ with a maximum of $n$ users,
Bil{\`o}~\cite{bilo2012unifying} proposes the following linear program for 
computing an upper bound $\opt{\gamma}$ on the price of anarchy:
\begin{equation} \label{eq:bilolp}
\begin{aligned}
    \opt{\gamma} =\> \underset{\kappa\in\bb{R}_{\geq 0}, \gamma\in\bb{R}}{\text{maximize}} \quad & \gamma \\
    \text{subject to:} \quad & \gamma y^2 - x^2 + \kappa [x^2-(x+1)y] \geq 0, 
    \quad \forall x,y \in \{0,1,\dots,n\}.
\end{aligned}
\end{equation}
We observe that solving the linear programs in Equations~\eqref{eq:characterize_poa_lp} 
and~\eqref{eq:bilolp} for the set of affine congestion games $\cal{G}$ with 
$n\leq 2$ users yields ${\rm PoA}(\cal{G})= 2$ and ${\rm PoA}(\cal{G})\leq 2.5$, 
respectively.%
\footnote{
    One can verify that the solution to the linear program in 
    Equation~\eqref{eq:characterize_poa_lp} is $(\opt{\nu},\opt{\rho})=(0.5,0.5)$, 
    while the solution to the linear program in Equation~\eqref{eq:bilolp} is 
    $(\opt{\kappa},\opt{\gamma})=(1.5,2.5)$.
}
It then holds that upper bounds on the price of anarchy derived from the tractable 
linear program in Reference~\cite{bilo2012unifying} are not tight, as 
${\rm PoA}(\cal{G})=2<2.5$ in this example.
\end{example}

Observe that the linear program in Equation~\eqref{eq:characterize_poa_lp} closely 
resembles the linear program in Equation~\eqref{eq:bilolp}.
In fact, these two linear programs are identical in structure as they are both 
tractable reductions of the linear program in Equation~\eqref{eq:gsmooth_lp}.
They only differ in the parameterization of the constraint set.
In this respect, the game parameterization we identify in this work is critical in retaining tightness of the generalized price of anarchy using an extremely modest number of constraints.
In contrast, though the parameterization used in \cite{bilo2012unifying} 
has a comparable number of constraints, we observed in \cref{ex:bilo_loose} that it 
loses tightness.


\subsection{Optimizing the price of anarchy} \label{sec:optimal_incentive_design}
The previous section focused on how to characterize the price of anarchy in any set 
of generalized congestion games.  
In this section, we shift our focus to the derivation of cost generating 
functions that optimize the price of anarchy.  
That is, given a set of resource cost functions $C^1, \dots, C^m$, 
what is the corresponding set of cost generating functions $F^1, \dots, F^m$ 
that minimizes the resulting price of anarchy $\poa(\gee)$? 
Recall from the introduction that this line of questioning is relevant to 
the problem of incentive design given in Section~\ref{sec:motivating_example}, 
when the price of anarchy is the performance bound of interest.

The following theorem provides a tractable and scalable methodology for computing
the set of cost generating functions that minimize the price of anarchy.

\begin{theorem}\label{thm:optimize_poa}
Let $C^1, \dots, C^m$ denote a set of resource cost functions defined for $n$ users,
and let $(\optj{F}{j}, \optj{\rho}{j})$, $j=1,\dots,m$, be solutions to 
the following $m$ linear programs:
\begin{equation}\label{eq:opt_distr_rule}
    \begin{aligned}
        \underset{F \in \mathbb{R}^n, \rho \in \mathbb{R}}{\text{maximize}} \quad & \rho \\
        \text{subject to:} \quad & C^j(y) - \rho C^j(x) + (x-z)F(x) - (y-z)F(x+1) \geq 0, \quad \forall (x,y,z) \in \IR(n).
    \end{aligned}
\end{equation}
Then the cost generating functions $\optj{F}{1}, \dots, \optj{F}{m}$ minimize
the price of anarchy and the price of anarchy corresponding to basis function 
pairs $\{ C^j, \optj{F}{j} \}$, $j=1,\dots,m$, satisfies 
\[ \poa(\gee) = \max_{j \in \{1,\dots,m\}} \frac{1}{\optj{\rho}{j}}. \]
\end{theorem}

\cref{thm:optimize_poa} states that we can derive cost generating functions 
$\optj{F}{1}, \dots, \optj{F}{m}$ that minimize the price of anarchy by solving $m$ 
independent linear progams, where each $\optj{F}{j}$ can be derived using only 
information about the corresponding resource cost function $C^j$. 
Accordingly, the price of anarchy of this optimized system corresponds to the worst 
price of anarchy associated with any single pair $\{C^j,\optj{F}{j}\}$, i.e., 
\[ \poa(\gee) = \max_{j \in \{1, \dots,m\}} \poa(\gee^j), \] 
where $\gee^j \subseteq \gee$ represents the set of generalized congestion games 
induced by $n$ and the basis function pair $\{C^j, \optj{F}{j}\}$.
Observe that this statement is not true in general for an arbitrary set of basis 
function pairs, i.e., there exist sets of basis function pairs $\{C^j, F^j\}$, 
$j=1,\dots,m$, such that\footnote{
    For example, consider the set of generalized congestion games $\gee$ induced by $n = 3$, 
    and $\{ \{C^1,F^1\}, \{C^2,F^2\} \}$, where $\{C^1(k), F^1(k) \} = \{ k^2, k \}$ and 
    $\{C^2,F^2\} = \{ k, k\}$ for all $k = 1, \dots, n$.
    Using the linear program in Equation~\eqref{eq:characterize_poa_lp}, we get 
    $\poa( \gee^1 ) ) = 2.5$, $\poa( \gee^2 ) ) = 2.0$, 
    and $\poa( \gee ) = 2.6$.
    For this particular choice of $\gee$, observe that 
    $\poa( \gee ) > 
        \max_{j \in \{1, \dots, m\}} \poa( \gee^j )$.
}
\[ \poa(\gee) > \max_{j\in\{1,\dots,m\}} \poa(\gee^j). \]
However, when we restrict our attention to optimal cost generating functions 
for each $C^j$, the above strict inequality holds with equality.
This is the key observation in the proof of \cref{thm:optimize_poa}.


\begin{proof}[Proof of \cref{thm:optimize_poa}.]

For each $j \in \{1,\dots,m\}$,
the function $\optj{F}{j}$ maximizes $\optj{\rho}{j}$ by the 
following reasoning: 
For each resource cost function $C^j$, we wish to find the function $\optj{F}{j}$ that 
maximizes $\rho$ in Equation~\eqref{eq:characterize_poa_lp}. 
Finding such a function is equivalent to finding the solution to
\begin{align*}
    (\optj{F}{j}, \optj{\nu}{j}, \optj{\rho}{j}) 
        \in & \argmax_{F \in \mathbb{R}^n,\nu \in \mathbb{R}_{\geq 0}, \rho \in \mathbb{R}} 
            \quad \rho\\
    \text{s.t.} & \quad C^j(y) - \rho C^j(x) + \nu [(x-z)F(x) - (y-z)F(x+1)] \geq 0, 
            \quad \forall (x,y,z) \in \IR(n).
\end{align*}
It is important to note that an optimal function $\optj{F}{j}$ must exist since the above program
is feasible for $F^j(k)=0$, $k=1,\dots,n$, $\nu=1$ and $\rho\leq \min_{x,y} C^j(y)/C^j(x)$,
and is bounded since any pair $\{C^j,\optj{F}{j}\}$ generates a set of games $\cal{G}^j$ so 
$\optj{\rho}{j}=1/{\rm PoA}(\cal{G}^j) \in [0,1]$ must hold by \cref{thm:gs_gcg}.

To obtain a linear program, we combine the decision variables
$\nu$ and $F$ in $\tilde{F}(k) := \nu F(k)$ to get
\begin{align*}
    (\tilde{F}^j_\mathrm{opt}, \tilde{\rho}^j_\mathrm{opt}) \in 
        \argmax_{F \in \mathbb{R}^n, \rho \in \mathbb{R}} \quad & \rho \\
    \text{s.t.} \quad 
        & C^j(y) - \rho C^j(x) + (x-z)F(x) - (y-z)F(x+1) \geq 0, 
            \quad \forall (x,y,z) \in \IR(n).
\end{align*}
Note that $\tilde{F}^j_\mathrm{opt} \in \mathbb{R}^n$ must be feasible as 
$\optj{\tilde{F}}{j}(k) = \optj{\nu}{j} \optj{F}{j}(k)$, and we know that 
$\optj{F}{j} \in \mathbb{R}^n$ exists.
Further, $\optj{\tilde{\rho}}{j} = \optj{\rho}{j}$, 
as equilibrium conditions are invariant to scaling of $F$.

For the set of generalized congestion games $\gee$ induced by $n$ 
and basis function pairs $\{C^j, \optj{\tilde F}{j}\}$, $j=1,\dots,m$,
and the set of games $\gee^j$ induced by $n$ and the basis function 
pair $\{C^j, \optj{\tilde F}{j}\}$, it holds that
$ \poa(\gee) \geq \max_{j \in \{ 1, \dots, m \}} \poa( \gee^j )$.
We conclude by proving that the converse also holds, i.e.,
\begin{equation*}
    \poa( \gee ) \leq \max_{j \in \{ 1, \dots, m \}} \poa( \gee^j ).
\end{equation*} 
Simply note that the values $(\nu,\rho)=(1,\opt{\rho})$ are feasible in the linear program 
in Equation~\eqref{eq:characterize_poa_lp} for the function pairs $\{C^j,\optj{\tilde F}{j}\}$, $j=1,\dots,m$,
where $\opt{\rho} := \min_j \optj{\rho}{j}$.
This implies that $\poa( \gee ) \leq 1/\opt{\rho}$. 
Observing that $1/\opt{\rho} = \max_{j \in \{ 1, \dots, m \}} \poa( \gee^j )$ concludes the proof.

\end{proof}


\section{Generalized Smoothness in Welfare Maximization Games} \label{sec:welfare_maximization_games}

Although the primary focus of this paper is on cost minimization settings, many of the results that
we obtain can be analogously derived for welfare maximization problems.
A welfare maximization problem consists of a set $N = \{1, \dots, n\}$ of users, where each user 
$i \in N$ is associated with a finite action set $\aset_i$.
The global objective is to \emph{maximize} the system's welfare, which is measured by the welfare 
function $W:\aset \to \mathbb{R}_{> 0}$, i.e. we wish to find the allocation $\opt{a} \in \aset$, such 
that $\opt{a} \in \argmax_{a \in \aset} W(a)$.
As with cost minimization problems, we consider a game-theoretic model where each user $i\in N$ is 
associated with a utility function $U_i:\aset \to \mathbb{R}$, which it uses to evaluate its own 
actions against the collective actions of the other users. 
A \emph{welfare maximization game} is a tuple $G = (N,\aset,W,\{U_i\})$.
Given a welfare maximization game $G$, a pure Nash equilibrium is defined as an allocation 
$\nash{a} \in \aset$ such that $U_i(\nash{a}) \geq U_i(a_i, \nash{a}_{-i})$ for all $a \in \aset_i$, and all 
$i \in N$.
The price of anarchy in welfare maximization games is defined similarly to Equation~\eqref{eq:poa} and 
Equation~\eqref{eq:poa_class},%
\footnote{%
    For consistency with the previous sections, we opt to define the price of anarchy in 
    welfare maximization games as the ratio between the welfare at the optimal allocation and the 
    system welfare at the worst performing Nash equilibrium, in contrast with previous works
    \citep{gairing2009covering,roughgarden2015intrinsic}. 
    This is achieved by inverting the ratio, i.e., defining the price of anarchy as the worst case ratio 
    between the welfare at optimum, and the welfare at the equilibria in $\mathrm{NE}(G)$. 
    By adopting this formalism, \mbox{we retain the overall objective of \emph{minimizing} the system's 
    price of anarchy.}
}
\begin{align*}%
    \poa(G) := \frac{\max_{a \in \aset} W(a)}{\min_{a \in \mathrm{NE}(G)} W(a)} \geq 1, \quad
    \poa(\cg) := \sup_{G \in \cg} \poa(G) \geq 1,
\end{align*}
where a lower value of the price of anarchy corresponds to an improvement in performance. 
%


%
\subsection{Generalized smoothness in welfare maximization games}
We begin with the definition of generalized smoothness in welfare maximization games and then
provide the analogue of \cref{thm:generalized_smoothness}.
\begin{definition} \label{def:generalized_smoothness_wm}
The welfare maximization game $G$ is ($\lambda,\mu$)-generalized smooth if, for any two allocations 
$a, a' \in \aset$, there exist $\lambda > 0$ and $\mu > -1$ satisfying,
\begin{equation} \label{eq:generalized_smoothness_wm}
    \sum^n_{i=1} U_i(a_i', a_{-i}) - \sum^n_{i=1} U_i(a) + W(a) \geq \lambda W(a') - \mu W(a).
\end{equation}
\end{definition}
\begin{proposition} \label{thm:welfare-maximization}
    The price of anarchy of a ($\lambda,\mu$)-generalized smooth welfare maximization game $G$ is 
    upper bounded as,
    \[ \poa(G) \leq \frac{1+\mu}{\lambda}. \]
\end{proposition}

We define the generalized price of anarchy of a set of welfare maximization games $\cg$ as 
\begin{equation} \label{eq:gpoa_wm}
    \gpoa(\cg) := \inf_{\lambda > 0, \mu > -1} \left\{ \frac{1+\mu}{\lambda} 
            \text{ s.t. Equation~\eqref{eq:generalized_smoothness_wm}  holds } \forall G \in \cg \right\}.
\end{equation}
As with cost minimization games, all efficiency guarantees also extend to average coarse-correlated 
equilibria (as in Observation~\#3) and there are also provable advantages of generalized smoothness 
over the original smoothness framework in terms of characterizing price of anarchy bounds (as in 
\cref{thm:generalized_smoothness}).  
We do not explicitly state or prove these parallel results to avoid redundancy.


\subsection{Generalized smoothness in distributed welfare games}
In this section, we consider \emph{distributed welfare games}~\cite{marden2013distributed} as
described in \cref{sec:motivating_example}, which are the welfare maximization analogue to 
generalized congestion games.  
Games in this class feature a set of users $N=\{1,\dots,n\}$ and a finite set of 
resources $\resset$.
The system welfare and user utility functions are defined as
\begin{align*}
    W(a) = \sum_{r \in \resset} W_r(|a|_r), 
    \qquad U_i(a) = \sum_{r \in a_i} F_r(|a|_r),
\end{align*}
where, for each $r \in \ree$, $W_r : \{ 0, 1, \dots, n \} \to \mathbb{R}_{\geq 0}$ and 
$F_r : \{ 1, \dots, n \} \to \mathbb{R}_{\geq 0}$ are the resource welfare function and 
utility generating function, respectively. 
For the remainder of this section, given basis function pairs $\{W^j, F^j\}$, $j=1,\dots,m$,
we define the set of local welfare maximization games $\gee$
in the same fashion as for generalized congestion games given in \cref{sec:slcsg}. 
Distributed welfare games have been used to model several problems of interest as in
\cite{barman2020tight,chandan2021tractable,gairing2009covering,kleinberg2011mechanisms,marden2013distributed}.

The following theorem provides the analagous results derived for generalized congestion games 
to the domain of distributed welfare games. 
As before, we define $W^j(0)=F^j(0)=F^j(n+1)=0$, for $j=1,\dots,m$, for ease of notation.  
\cref{thm:gs_lwm} is stated without proof as the reasoning follows almost identically 
to the proofs of \cref{thm:gs_gcg,thm:optimize_poa}.

\begin{theorem} \label{thm:gs_lwm}
Let $\cal{G}$ denote the set of all distributed welfare games with a maximum of $n$ users 
generated from basis function pairs $\{ W^j, F^j \}$, $j=1,\dots,m$. 
The following statements hold true:
\begin{enumerate}
    \item [(i)] The price of anarchy and the generalized price of anarchy satisfy 
\( \poa( \gee ) = \gpoa( \gee ). \)
\item[(ii)] Let $\opt{\rho}$ be the value of the following linear program:
\begin{equation} \label{linprog:characterize_poa_wm}
\begin{aligned}
    \opt{\rho} = \> & 
        \min_{\nu \in \mathbb{R}_{\geq 0}, \rho \in \mathbb{R}} \quad \rho \\
    & \text{s.t.} \quad W^j(y) - \rho W^j(x) + \nu\left[ (x-z)F^j(x)-(y-z)F^j(x+1) \right] \leq 0 \\
    & \hspace{150pt} \forall j = 1,\dots,m, \quad \forall (x,y,z) \in \IR(n),
\end{aligned}
\end{equation}
Then, it holds that $\poa( \gee ) = \opt{\rho}$.
\item[(iii)] Let the parameters $(\optj{F}{j}, \optj{\rho}{j})$, $j=1,\dots,m$, 
be solutions to the following $m$ linear programs:
\begin{equation} \label{linprog:optimal_poa_wm}
\begin{aligned}
    ( \optj{F}{j}, \optj{\rho}{j} ) \in 
    & \argmin_{F \in \mathbb{R}^n, \rho \in \mathbb{R}} \> \rho \quad \text{subject to:}\\
    & W^j(y) - \rho W^j(x) + (x-z)F(x) - (y-z)F(x+1) \leq 0, 
            \quad \forall (x,y,z) \in \IR(n).
\end{aligned}
\end{equation}
Then, the utility generating functions $\optj{F}{1}, \dots, \optj{F}{m}$ minimize the price 
of anarchy, and the price of anarchy corresponding to basis function pairs $\{W^j, \optj{F}{j}$\}, 
$j=1,\dots,m$, satisfies  
\begin{equation*}
    \poa( \gee ) = \max_{j \in \{1, \dots, m\}} \optj{\rho}{j}.
\end{equation*}
\end{enumerate}
\end{theorem}


\section{Illustrative Examples} \label{sec:illustrative_examples}

In the introduction, we motivated our study by considering two seemingly distinct 
problems: incentive design in congestion games and utility design in distributed welfare games.
In this section, we utilize these same classes of problems (among others) to demonstrate the breadth of our approach.
For an in-depth study discussing the application of the machinery derived here to the design 
of incentives in congestion games, we refer to Paccagnan et al.~\cite{paccagnan2019incentivizing}.

\subsection{Price of anarchy in congestion games and their variants}
\cref{thm:gs_gcg} allows to determine the exact price of anarchy for any game that can be cast as generalized congestion game. In this section we illustrate the applicability of this result by i) recovering/extending classical findings on the price of anarchy of congestion games, ii) computing the efficiency of marginal cost incentives, iii) providing novel price of anarchy results for \mbox{perception-parametrized congestion games.}

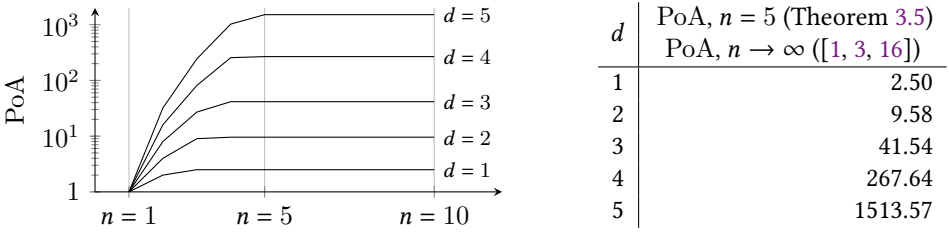
\begin{figure}[ht]
    {
    \begin{minipage}[b]{0.50\linewidth}
        \centering
        \begin{tikzpicture}
            \begin{semilogyaxis}
            [
            height=4cm,
            axis x line = bottom,
            axis y line = left,
            ylabel={${\rm PoA}$},
            xmin=0, xmax=12,
            ymin=1, ymax=2000,
            xtick={1,5,10},
            xticklabels={$n=1$, $n=5$, $n=10$},
            ytick={1,10,100,1000},
            yticklabels={$1$, $10^1$, $10^2$, $10^3$},
            xmajorgrids,
            legend pos=outer north east,
            ]
            \addplot[no marks] coordinates{
                (1,1) (2,2) (3,2.5) (10,2.5)
            }
            [pos segment=0,font=\footnotesize]
                node [right,pos=1] {$d=1$};
            \addplot[no marks] coordinates{
                (1,1) (2,4) (3,9) (4,9.5833) (10,9.5833)
            }
            [pos segment=0,font=\footnotesize]
                node [right,pos=1] {$d=2$};
            \addplot[no marks] coordinates{
                (1,1) (2,8) (3,27) (4,41.5357) (10,41.5357)
            }
            [pos segment=0,font=\footnotesize]
                node [right,pos=1] {$d=3$};
            \addplot[no marks] coordinates{
                (1,1) (2,16) (3,81) (4,256) (5,267.6432) (10,267.6432)
            }
            [pos segment=0,font=\footnotesize]
                node [right,pos=1] {$d=4$};
            \addplot[no marks] coordinates{
                (1,1) (2,32) (3,243) (4,1024) (5,1513.57) (10,1513.57)
            }
            [pos segment=0,font=\footnotesize]
                node [right,pos=1] {$d=5$};
            \legend{};
            \end{semilogyaxis}
        \end{tikzpicture}
    \end{minipage}%
    \begin{minipage}[b]{0.50\linewidth}
        \centering
        \begin{tabular}[b]{c|r} 
            \multirow{2}{*}{$d$} & \multicolumn{1}{c}{\,${\rm PoA}$, $n=5$ (\cref{thm:gs_gcg})} \\          
             & \multicolumn{1}{c}{${\rm PoA}$, $n\to\infty$~(\cite{aland2011exact,awerbuch2005price,christodoulou2005price})}\\
            \hline
            1 &     2.50  \\
            2 &     9.58 \\
            3 &    41.54  \\
            4 &   267.64  \\
            5 &  1513.57  \\
            \multicolumn{2}{c}{\vspace*{-3.5mm}}\\
        \end{tabular}
    \end{minipage}%
    }
    \caption{
   Evolution of the price of anarchy in polynomial congestion games of order $d=1,\dots,5$ as a function of the number of users (left).
    These values were obtained by solving the corresponding linear program in \cref{thm:gs_gcg}.
   Observe that the price of anarchy plateaus at $n=5$, matching the asymptotic bounds ($n\rightarrow \infty$) previously obtained in the literature \cite{aland2011exact,awerbuch2005price,christodoulou2005price}. This suggests that small instances are sufficient to produce highly inefficient equilibria.
    }\label{fig:congestion_poas}
        \end{figure}

\paragraph{i) Congestion games}
Congestion games constitute a subclass of problems to which \cref{thm:gs_gcg} applies. This follows readily upon letting $C_r(k) = k \cdot c_r(k)$, $F_r(k) = c_r(k)$ for $k=1,\dots,n$ and $r\in\cal{R}$ in Equations~\eqref{eq:gc_system_cost} and \eqref{eq:gc_user_cost}, where $c_r(\cdot)$ describes the original resource congestion.
Hence, we are able to compute their price of anarchy by simply solving the linear program in \cref{thm:gs_gcg}. As a special case, we recover well-known price of anarchy results \cite{aland2011exact,awerbuch2005price,christodoulou2005price} for polynomial congestion games of maximum degree $d$, i.e., congestion games where the resource congestion is obtained by non-negative linear combinations of monomials $1,x,\dots,x^d$.
Although the bounds provided in these works are exact (for large $n$), the authors had to combine traditional smoothness arguments with nontrivial worst-case game constructions.\footnote{Limited to this settings, the smoothness and generalized smoothness inequalities coincide since the system cost equals the sum of the users' costs when no incentives are employed.}
In contrast, the linear program in \cref{thm:gs_gcg} provides exact price of anarchy values for all $n$, can be solved efficiently (featuring only two decision variables and $(d+1)\cal{O}(n^2)$ constraints), and does not require ad-hoc worst-case constructions. 
We solve such program as a function of the number of users $n$ and the maximum degree $d$, reporting the results in~\cref{fig:congestion_poas}. To the best of our knowledge, this is the first characterization of the dependence of the price of anarchy in polynomial congestion games on the number of users $n$. Remarkably, we note (table in~\cref{fig:congestion_poas}), that the price of anarchy values for $n=5$ users exactly match their corresponding asymptotic values ($n\rightarrow \infty$) from \cite{aland2011exact,awerbuch2005price,christodoulou2005price}, suggesting that very small instances are sufficient to produce highly inefficient equilibria.

Finally, we remark that the machinery developed here can be used to characterize the price of anarchy for a variety of congestion functions often employed in the literature.
This includes the well-studied Bureau of Public Roads (BPR) function~\cite{united1964traffic} where 
$c_r(x) = T_r \cdot \Big[ 1+0.15\cdot\Big(\frac{x}{K_r}\Big)^4 \Big]$, and $T_r\geq 0$, $K_r\in\bb{N}_{\geq 1}$ are the free flow congestion and capacity of road $r$.
Solving the corresponding linear program in \cref{thm:gs_gcg}, one obtains a price of anarchy of approximately $36.09$ for $n=50$ users and $K_r\in\{1,\dots,50\}$.
This highlights that, although BPR functions are polynomials of order $d=4$, their special structure allows significant reductions in the price of anarchy \mbox{(from $267.64$ to $36.09$).}

\paragraph{ii) Marginal cost incentives in congestion games}
Marginal cost incentives have been repeatedly proposed to improve the performance of Nash equilibria in congestion games, e.g., \cite{maille2009eliciting,pigou1920economics}. In the nonatomic variant of this model (whereby users are treated as divisible entities), these incentives guarantee optimal equilibrium efficiency, i.e., their price of anarchy is exactly $1$. In the classical atomic setting, marginal cost incentives take the form
\begin{equation*}
    \tau_r(k) = (k-1)[c_r(k) - c_r(k-1)],
\end{equation*}
allowing for the deployment of our framework to compute their efficiency. This follows readily upon letting $C_r(k) = k\cdot c_r(k)$ and $F_r(k) = k\cdot c_r(k) - (k-1)\cdot c_r(k-1)$ for all $k$ and $r$ in Equations~\eqref{eq:gc_system_cost} and \eqref{eq:gc_user_cost}. Thus, using the linear program in \cref{thm:gs_gcg}, we compute the corresponding price of anarchy for polynomial congestion games of order $d=1,\dots,5$ with $n=100$ (Column~3 in \cref{tab:poa_rebates}).
Perhaps surprisingly, while marginal cost incentives promote optimal performance in the nonatomic settings, their use in the atomic model significantly deteriorates the system's efficiency, with a price of anarchy greater than that experienced when no incentives are used (Columns 2, 3 in \cref{tab:poa_rebates}).

\paragraph{iii) Perception-parametrized congestion games}
The \emph{perception-parameterized} congestion game model was proposed by Kleer and Sch{\"a}fer~\cite{kleer2019tight} to unify the notions of risk sensitivity~\cite{caragiannis2012computing,piliouras2016risk}, and altruism~\cite{caragiannis2010impact,chen2014altruism} in affine congestion games.
In this model, the system and user costs are
\[    C(a) = \sum_{r\in\cal{R}} |a|_r \cdot c_r(1+\sigma(|a|_r-1)), \qquad J_i(a) = \sum_{r\in a_i} c_r(1+\gamma(|a|_r-1)), \]
where $\sigma, \gamma \geq 0$ are fixed parameters and $c_r(x)$ 
is an affine function.
Among other parameterizations, $\sigma=\gamma=1$ models ``classical'' congestion games as in \cref{ex:congestion_games} and $\sigma=1$, $\gamma\geq 1$ models congestion games with altruistic users, whereas $\sigma=\gamma\geq 0$ describes congestion games in which each user $i\in N$ participates in the game with probability $p_i=\sigma=\gamma$ \cite{cominetti2019price}. 
Note that, for given $\sigma, \gamma \geq 0$, the corresponding class of perception-parameterized congestion games is covered by our framework. To see this, it suffices to set 
$C_r(k)=k\cdot c_r(1+\sigma(k-1))$, $F_r(k)=c_r(1+\gamma(k-1))$ for all $k$ and $r$ in Equations~\eqref{eq:gc_system_cost} and \eqref{eq:gc_user_cost} to cover the case of affine resource costs as well as more general cases.
Thus, evaluating the price of anarchy of perception-parameterized congestion games -- which remains an open problem even in the affine case -- is equivalent to solving the linear program in \cref{thm:gs_gcg}.
In \cref{fig:perception}, we plot the solution for the affine case, $n=20$ users and $\sigma,\gamma\in[0,2]$. Not only do we recover the exact asymptotic bounds of \cite{kleer2019tight} where applicable (region enclosed by white line), but we also provide a complete characterization of the price of anarchy for all $\sigma,\gamma\in[0,2]$.

\pgfplotsset{every tick label/.append style={font=\tiny}}

\begin{figure}[h]{
    \centering
    \begin{tikzpicture}[
        scale=0.75
    ]
    \begin{axis}[xmin=0,xmax=2,xtick={0,0.25,0.5,...,2},ymin=1,ymax=10,xlabel={$\gamma$},ylabel={$\poa$},width=8cm,height=5cm]
        \addlegendentry{$\sigma=0.5$};
        \addlegendentry{$\sigma=1.0$};
        \addlegendentry{$\sigma=1.5$};
        \addlegendentry{$\sigma=2.0$};
        \addplot[blue,thick,smooth,no markers] table[x={x},y={0.5}]{poa.dat};
        \addplot[red,thick,smooth,no markers] table[x={x},y={1}]{poa.dat};
        \addplot[brown,thick,smooth,no markers] table[x={x},y={1.5}]{poa.dat};
        \addplot[thick,smooth,no markers] table[x={x},y={2}]{poa.dat}; 
    \end{axis}
    \node (2dplot) at (-3.5,0) {};
    \node[inner sep=0pt] (poa) at (-5.1,1.7)
        {\includegraphics[width=7.3cm]{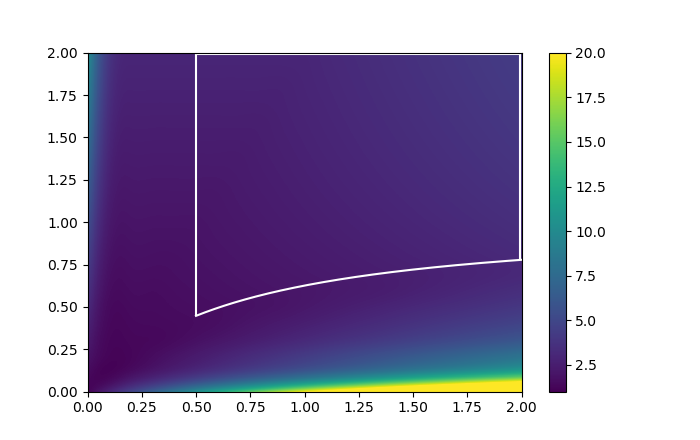}};
    \node[rotate=90] () at (-10,1.7) {$\gamma$};
    \node[align = center, below = -0.25cm of poa] {$\sigma$};
    \end{tikzpicture}
    }
    \caption{Exact price of anarchy for perception-parameterized affine congestion games with $n=20$ users and $\sigma,\gamma \in [0,2]$, computed via \cref{thm:gs_gcg} (left). Corresponding values for fixed $\sigma\in\{0.5,1,1.5,2\}$ (right). Kleer and Sch{\"a}fer~\cite{kleer2019tight} give asymptotic values limited to the region enclosed in the white perimeter, which we recover exactly and generalize.}
    \label{fig:perception}
\end{figure}
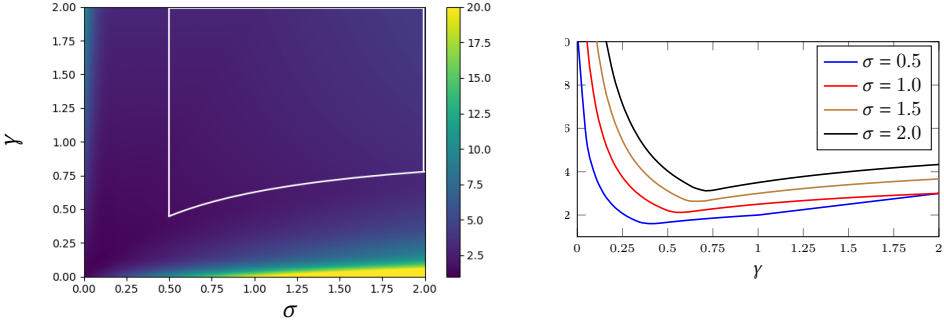

\subsection{Optimal local incentives in congestion games} 

In this section we consider \emph{local} incentives, i.e, incentives that map each resource $r$ of the game to an incentive function $\tau_r$ by leveraging solely information on the corresponding congestion function $c_r$.
Whilst previous works also consider incentives that utilize global information, e.g., \cite{bilo2016dynamic, caragiannis2010taxes}, incentives based solely on local information have a number of advantages including limited informational requirements, scalability, ability to accommodate resources that are dynamically added or removed, and robustness against a number of variations.
To ease presentation, we will refer to local incentives simply as incentives.

As illustrated in the previous section, \cref{thm:gs_gcg} allows us to evaluate the price of anarchy of commonly studied classes of games, e.g., congestion games and generalization thereof. In contrast, we observe that \cref{thm:optimize_poa} does \emph{not} directly provide a machinery for the design of optimal incentives. 
To see this observe that, while \cref{thm:optimize_poa} allows to determine the best \emph{linear} incentive, optimal incentives might very well not satisfy this structural property.%
\footnote{Given a set of congestion games where each resource cost is obtained by the linear combination $c_r(k)=\sum^m_{j=1} \alpha^j_r \cdot c^j(k)$, we say that an incentive $T$ is \emph{linear} if it satisfies $T(c_r)=\sum^m_{j=1} \alpha^j_r \cdot T(c^j)$ (i.e., if it is obtained by a linear combination of the incentives $\{T(c^j)\}_{j=1}^m$ using the \emph{same} coefficients that define $c_r$).} 
Surprisingly, Paccagnan et al. \cite{paccagnan2019incentivizing} recently showed that the best linear incentive is \emph{optimal} (i.e., its performance can not be improved, even by a nonlinear incentive). %
Building upon this result, we are then guaranteed that the incentives derived in \cref{thm:optimize_poa} are the best possible.
Thus, we solve the linear program derived in \cref{thm:optimize_poa} and report the optimal values of the price of anarchy for polynomial congestion games of degree $d= 1,\dots,5$ and $n = 100$ in Column~5 of \cref{tab:poa_rebates}. We observe that the achieved price of anarchy is significantly lower (better) than the setting without incentives (Column~2).

We conclude the section by highlighting that our framework can also accommodate commonly-studied constraints on the set of admissible incentives. For example, fixed incentives (i.e., incentives that are constant in the congestion) can be studied by imposing $\tau_r(k)=\tau_r$, which corresponds to substituting $F^j(k)=c^j(k)+\tau^j$, into the linear program in \cref{thm:gs_gcg}. Including $\sigma^j=\nu\cdot\tau^j$ as decision variables, we obtain a linear program with $m+2$ decision variables and $\cal{O}(mn^2)$ constraints for computing the optimal fixed incentives. We report the corresponding optimal price of anarchy for polynomial congestion games of degree $d=1,\dots,5$ in Column~6 of \cref{tab:poa_rebates}, and observe that such simple incentives already provide a good improvement upon the setting without incentives.

\begin{table}[h]
\label{tab:poa_rebates}
\caption{\emph{Price of anarchy in polynomial congestion games.}
Price of anarchy for polynomial congestion games with degree $d = 1, \dots, 5$ and $n = 100$.
The second column contains the asymptotic values ($n\rightarrow\infty$) without incentives 
\citep{aland2011exact,awerbuch2005price,christodoulou2005price}.
In the third column, we report the values corresponding to the use of marginal cost incentives (computed through \cref{thm:gs_gcg}).
The fourth and fifth columns feature optimal price of anarchy values for general and fixed local incentives, respectively (computed through \cref{thm:gs_gcg,thm:optimize_poa}).
}
\begin{tabular*}{\textwidth}{ c | @{\extracolsep{\fill}} r | r | r | r }
    \multirow{2}{*}{$d$} & \multirow{2}{*}{No Incentive} & \multirow{2}{*}{Marginal Cost} &
    \multirow{2}{*}{Optimal Local Incentives} & 
    \multirow{2}{*}{Optimal Fixed Incentives}\\
    &&&& \\
    \hline
        1 &     2.50 &     3.00 &   2.012 &    2.15\\
        2 &     9.58 &    13.00 &   5.101 &    5.33\\
        3 &    41.54 &    57.36 &  15.551 &   18.36\\
        4 &   267.64 &   391.00 &  55.452 &   89.41\\
        5 &  1513.57 &  2124.21 & 220.401 &  469.74
\end{tabular*}
\end{table}

\subsection{Optimal utility design in distributed welfare games}
The preceding discussion showcases how the machinery we developed can be used to compute and optimize the price of anarchy in well-studied classes of cost minimization games. While a similar approach can be followed also for welfare maximization problems (and in fact, recent results build on top of this work to derive state-of-the-art approximation algorithms with explicit guarantees \cite{barman2020tight,chandan2021tractable,paccagnan2019incentivizing,vijayalakshmi2020improving}), we purposely decide to take a different perspective in this section. Specifically, we aim at demonstrating the \emph{robustness} of the proposed approach and the \emph{quality} of the corresponding results.

Toward this goal, rather than fixing a specific set of welfare functions, we consider a general setting, whereby $W_r:\{1,\dots,n\}\to\bb{R}$ merely satisfies two properties: non-decreasingness, i.e., $W_r(k+1)\geq W_r(k)$, and concavity (or diminishing returns property), i.e., $W_r(k+1)-W_r(k)\leq W_r(k)-W_r(k-1)$, for all $k$ and $r$. Observe that these properties are commonly encountered in  application areas including vehicle-target assignment problems \cite{chopra2017distributed,murphey2000target}, multiwinner elections \cite{dudycz2020tight} and sensor coverage \cite{gairing2009covering,zhu2013distributed}. For any given welfare function satisfying non-decreasingness and concavity, we compare the performance (price of anarchy) obtained by \emph{optimal utilities} against that of the commonly-advocated-for \emph{identical interest} design, whereby each user's utility coincides with the system welfare, i.e.,  $U_i(a)=W(a)$. 
We do so for $10^5$ unique resource welfare functions $W_r$, randomly generated by sorting 10 independently values from a uniform distribution over $[0,1]$ from largest to smallest and setting $W_r(k)$ to be the sum over the first $k$ sorted values. Nondecreasingness and concavity of $W_r$ follow readily. For each generated resource welfare, we determine the corresponding optimal price of anarchy and the price of anarchy of the identical interest design through the solution of the linear programs in \eqref{linprog:optimal_poa_wm} and \eqref{linprog:characterize_poa_wm}, respectively.\footnote{%
Instead of determining the price of anarchy of the identical interest utilities directly, we can compute the price of anarchy of the marginal contribution utilities (taking the form $U_i(a)=W(a)-W(\emptyset_i,a_{-i})$), as these two values coincide. To see this note that the underlying set of Nash equilibria remains the same under either utility as, for any $a,a'\in\cal{A}$, since $W(a)-W(a_i',a_{-i}) \geq 0 \iff W(a)-W(\emptyset_i,a_{-i}) - [W(a'_i,a_{-i})-W(\emptyset_i,a_{-i})] \geq 0$. Here, $W(\emptyset_i,a_{-i})$ denotes the system welfare when user $i$ selects no action and the remaining users select their action in $a$.}
The left panel in \cref{fig:hist} depicts the resulting empirical distribution of the price of anarchy values, while the right reports the ratio between the price of anarchy in the identical interest and optimal settings (this ratio is never lower than one, as expected). %
We conclude by observing that, whilst the identical interest design may initially appear to be an intuitive and appealing option, \cref{fig:hist} clearly highlights that strictly better performance can be readily achieved using the machinery developed here.

\pgfplotsset{every tick label/.append style={font=\tiny}}

\begin{figure}[h]{
    \centering
    \hspace{-6.5cm}
    \begin{tikzpicture}[scale=0.75]
    \begin{axis}[
    width=8cm,height=5cm,
    ymin=0, ymax=0.3,
    minor y tick num = 3,
    ytick={0,0.1,0.2,0.3},
    yticklabels={0\%,10\%,20\%,30\%},
    area style,
    ylabel={Percentile},
    xlabel={Price of Anarchy},
    ]
        \addplot+[ybar interval,const plot,fill,draw] coordinates {    
            (1.02985852e+00, 7.00000000e-05)
            (1.04984415e+00, 9.70000000e-04)
            (1.06982978e+00, 4.49000000e-03)
            (1.08981541e+00, 1.31300000e-02)
            (1.10980104e+00, 2.94900000e-02)
            (1.12978668e+00, 5.00400000e-02)
            (1.14977231e+00, 6.69700000e-02)
            (1.16975794e+00, 8.40400000e-02)
            (1.18974357e+00, 9.06600000e-02)
            (1.20972920e+00, 9.26200000e-02)
            (1.22971484e+00, 8.83200000e-02)
            (1.24970047e+00, 8.30100000e-02)
            (1.26968610e+00, 7.30700000e-02)
            (1.28967173e+00, 6.22500000e-02)
            (1.30965737e+00, 5.35100000e-02)
            (1.32964300e+00, 4.44400000e-02)
            (1.34962863e+00, 3.57800000e-02)
            (1.36961426e+00, 2.93100000e-02)
            (1.38959989e+00, 2.28200000e-02)
            (1.40958553e+00, 1.81200000e-02)
            (1.42957116e+00, 1.45600000e-02)
            (1.44955679e+00, 1.11000000e-02)
            (1.46954242e+00, 8.47000000e-03)
            (1.48952805e+00, 6.72000000e-03)
            (1.50951369e+00, 4.50000000e-03)
            (1.52949932e+00, 3.34000000e-03)
            (1.54948495e+00, 2.70000000e-03)
            (1.56947058e+00, 1.82000000e-03)
            (1.58945621e+00, 1.31000000e-03)
            (1.60944185e+00, 7.40000000e-04)
            (1.62942748e+00, 6.60000000e-04)
            (1.64941311e+00, 3.10000000e-04)
            (1.66939874e+00, 3.00000000e-04)
            (1.68938438e+00, 1.40000000e-04)
            (1.70937001e+00, 8.00000000e-05)
            (1.72935564e+00, 7.00000000e-05)
            (1.74934127e+00, 5.00000000e-05)
            (1.76932690e+00, 1.00000000e-05)
            (1.78931254e+00, 0.00000000e+00)
            (1.80929817e+00, 1.00000000e-05)};
        \addlegendentry{Identical interest};
        \addplot+[ybar interval,const plot,fill,opacity=0.75,draw] coordinates {
            (1,0.0)
            (1.01566923,1.8300e-3)
            (1.03532873,4.2640e-2)
            (1.05498823,1.7196e-1)
            (1.07464773,2.6377e-1)
            (1.09430723,2.3537e-1)
            (1.11396673,1.4802e-1)
            (1.13362623,7.5800e-2)
            (1.15328574,3.3340e-2)
            (1.17294524,1.4970e-2)
            (1.19260474,6.4300e-3)
            (1.21226424,3.1700e-3)
            (1.23192374,1.3300e-3)
            (1.25158324,7.1000e-4)
            (1.27124274,4.4000e-4)
            (1.29090224,9.0000e-5)
            (1.31056175,5.0000e-5)
            (1.33022125,5.0000e-5)
            (1.34988075,3.0000e-5)};
        \addlegendentry{Optimal utilities};
        \draw[black,thick=1mm] (axis cs:1.259,0)--(axis cs:1.259,8.30100000e-02) node [right] {\bf 1.259};
        \draw[black,thick=1mm] (axis cs:1.1,0)--(axis cs:1.1,2.3537e-1) node [right] {\bf 1.100};
    \end{axis}
    \hspace{6cm}
    \begin{axis}[width=8cm,height=5cm,
    ymin=0, ymax=0.12,
    minor y tick num = 3,
    ytick={0,0.04,0.08,0.12},
    yticklabels={0\%,4\%,8\%,12\%},
    area style,
    ylabel={Percentile},
    xlabel={Improvement Factor},
    ]
        \addplot+[ybar interval,const plot,fill] coordinates {    
            (1.0, 0.0)
            (1.01397038e+00, 5.10000000e-04)
            (1.02701911e+00, 4.67000000e-03)
            (1.04006783e+00, 1.77800000e-02)
            (1.05311656e+00, 4.05300000e-02)
            (1.06616528e+00, 6.56000000e-02)
            (1.07921401e+00, 8.41100000e-02)
            (1.09226273e+00, 9.46600000e-02)
            (1.10531145e+00, 9.64000000e-02)
            (1.11836018e+00, 9.10400000e-02)
            (1.13140890e+00, 8.44700000e-02)
            (1.14445763e+00, 7.49400000e-02)
            (1.15750635e+00, 6.37800000e-02)
            (1.17055508e+00, 5.58100000e-02)
            (1.18360380e+00, 4.45600000e-02)
            (1.19665253e+00, 3.75900000e-02)
            (1.20970125e+00, 2.93000000e-02)
            (1.22274998e+00, 2.40700000e-02)
            (1.23579870e+00, 1.99400000e-02)
            (1.24884742e+00, 1.58200000e-02)
            (1.26189615e+00, 1.21400000e-02)
            (1.27494487e+00, 9.84000000e-03)
            (1.28799360e+00, 7.88000000e-03)
            (1.30104232e+00, 6.10000000e-03)
            (1.31409105e+00, 4.95000000e-03)
            (1.32713977e+00, 3.71000000e-03)
            (1.34018850e+00, 2.64000000e-03)
            (1.35323722e+00, 2.19000000e-03)
            (1.36628594e+00, 1.58000000e-03)
            (1.37933467e+00, 9.90000000e-04)
            (1.39238339e+00, 7.80000000e-04)
            (1.40543212e+00, 4.20000000e-04)
            (1.41848084e+00, 4.60000000e-04)
            (1.43152957e+00, 3.40000000e-04)
            (1.44457829e+00, 1.30000000e-04)
            (1.45762702e+00, 1.10000000e-04)
            (1.47067574e+00, 7.00000000e-05)
            (1.48372446e+00, 3.00000000e-05)
            (1.49677319e+00, 1.00000000e-05)
            (1.50982191e+00, 3.00000000e-05)
            (1.52287064e+00, 2.00000000e-05)};
        \draw[black,thick=1mm] (axis cs:1.14378,0) -- (axis cs:1.14378,8.44700000e-02) node [right] {\bf 1.144};
    \end{axis}
    \end{tikzpicture}
    }
    \caption{
        \emph{Price of anarchy for identical interest and optimal utilities in distributed welfare games.}
        Left: Empirical distribution of the price of anarchy with optimal utilities vs. identical interest design.
        Right: Improvement factor when using the optimal design in place of the identical interest design.
        The mean of each distribution is indicated by a bisecting solid black line.
        Observe that the optimal utilities offer significant improvement over the identical 
        interest utilities, reducing the price of anarchy by a factor of approximately 1.144 on average.
        The values in the above charts were obtained using the linear programs in \cref{thm:gs_lwm}
        for nondecreasing, concave resource welfare functions.
        }
    \label{fig:hist}
\end{figure}
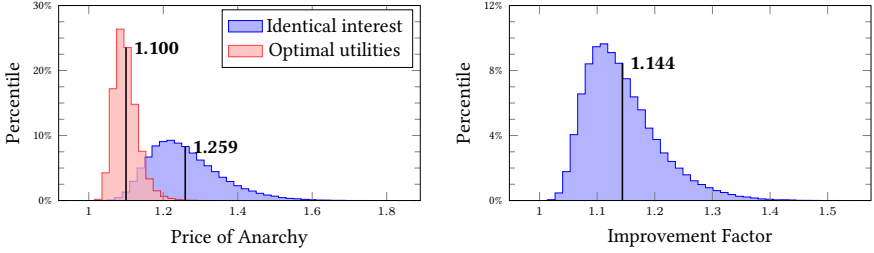


\section{Conclusions and Future Work} \label{sec:conclusion}
Though well-studied, the price of anarchy can still be difficult to compute as ad-hoc approaches are often needed. As a result, the design of incentives that optimize this metric is even more challenging, with only few results available. Motivated by this observation, our work provides a framework achieving two fundamental goals: to tightly characterize and optimize the price of anarchy through a computationally tracable approach.

Toward this end, we first introduced the notion of \emph{generalized smoothness}, which we showed always produces tighter or equal price of anarchy bounds compared to the original smoothness approach. We proved that such bounds are \emph{exact} for generalized congestion and local welfare maximization games, unlike those obtained through a simple smoothness argument. Additionally, we showed that the problems of computing and optimizing the price of anarchy can be posed (and solved) as tractable linear programs, when considering these broad problem classes.
Finally, we demonstrated the ease of applicability, strength and breadth of our approach by recovering and generalizing existing results on the computation of the price of anarchy, as well as by tackling the problems of incentive design in congestion games and utility design in distributed welfare games. In this regard, the list of illustrative example provided in \cref{sec:illustrative_examples} is certainly non-exhaustive.

Overall, we feel that the proposed approach has significant potential, especially since it can be used as a ``black box'' to compute and optimize the exact price of anarchy in many problems of interest. 
The linear programs derived here can be used, for example, as ``computational companions'' to support the analytical study of the price of anarchy, e.g, by providing evidence, or disproving certain conjectures. For this reason, we compliment our work with a software package that implements the techniques and linear programs derived here in the hope that they can be of help for new research to come.%
\footnote{\href{https://github.com/rahul-chandan/resalloc-poa}{https://github.com/rahul-chandan/resalloc-poa}}

We conclude observing that the price of anarchy represents but one of many metrics for measuring an algorithm's performance. Nevertheless, we believe that the techniques introduced here can be suitably extended to analyze different metrics (e.g., the price of stability) and to understand whether optimizing for the price of anarchy has any unintended consequences on them.

\begin{acks}
    A preliminary version of this work appeared in \cite{chandan2019when}. 
    This work is supported by ONR grants \#N00014-17-1-2060 and \#N00014-20-1-2359, 
    NSF grant \#ECCS-1638214, and SNSF grant \#P2EZP2\_181618.
\end{acks}

    \bibliographystyle{acm}
    \bibliography{references}
\end{document}